\documentclass[aps,prd,reprint,superscriptaddress,nofootinbib,longbibliography,floatfix]{revtex4-2}

\usepackage{amsmath}
\usepackage{amssymb}
\usepackage{graphicx}
\usepackage{hyperref}
\usepackage{xcolor}
\usepackage{bm}

\hypersetup{
    pdftitle={EMRI Dephasing from a Torsion-Inspired Near-Zone Kerr Deformation: A Phenomenological Study Motivated by Spin Matter},
    pdfauthor={Jingxu Wu, Liangyu Luo, Daniil Stepanenko, Jie Shi},
    colorlinks=true,
    linkcolor=blue,
    filecolor=magenta,
    urlcolor=cyan,
    citecolor=blue,
}

\begin{document}

\title{EMRI Dephasing from a Torsion-Inspired Near-Zone Kerr Deformation\\ A Phenomenological Study Motivated by Spin Matter}

\author{Jingxu Wu}
\thanks{These authors contributed equally to this work.}
\email[Corresponding author: ]{wuxj@my.msu.ru}
\affiliation{Faculty of Physics, Lomonosov Moscow State University, Moscow 119991, Russia}

\author{Liangyu Luo}
\affiliation{International School, I.M. Sechenov First Moscow State Medical University, Moscow 119048, Russia}

\author{Daniil Stepanenko}
\thanks{These authors contributed equally to this work.}
\email[Corresponding author: ]{dstepanenko@mi-ras.ru}
\affiliation{Steklov Mathematical Institute, Russian Academy of Sciences, 119333, Moscow, Russia}

\author{Jie Shi}
\affiliation{Faculty of Physics, Lomonosov Moscow State University, Moscow 119991, Russia}

\begin{abstract}
Extreme-mass-ratio inspirals (EMRIs) are sensitive to weak changes in a
massive compact object's exterior geometry. We quantify the circular
equatorial response of a prescribed near-zone Kerr deformation motivated
by spin matter. The metric is defined on a regular exterior domain;
its conservative orbital response is separated from the assumed radiation
law. For \(a/M=0.9\), \(\mu/M=10^{-5}\), and
\(\alpha/M^3=10^{-3}\), the ISCO radius increases by
\(3.6538\times10^{-3}M\), while its orbital frequency decreases by
\(3.9168\times10^{-4}/M\). At shared frequency endpoints corresponding
to Kerr radii \(8M\) and \(3M\), a frozen numerical Kerr flux gives
\(-435.69\) rad of conservative dephasing, compared with \(-460.75\) rad
for a quadrupole baseline. A Johannsen comparison matched in its leading
\(tt\) correction gives \(-64.87\) rad under the same flux, showing
sensitivity to the full metric prescription. Local timescale diagnostics
identify the failure of an adiabatic extrapolation toward the ISCO.
In the spherical Einstein--Cartan (EC) sector, we construct a six-state
Kramers-bundle spin representation on a Fermi--Walker-transported comoving
tetrad that satisfies four-dimensional Mathisson--Papapetrou--Dixon transport
and direct coordinate contraction $\tfrac12 s_{\mu\nu}s^{\mu\nu}=\chi_s n^2$,
showing explicitly that while a single polarized branch breaks spherical
isotropy, the local six-state coarse-grained ensemble closes the isotropic
effective stress tensor with $\langle\mathcal S^2\rangle=\chi_s n^2$ and
$\delta\langle\mathcal S^2\rangle=2\chi_s n\delta n$. On the nonlinear finite
cloud with a vacuum exterior, we solve the simultaneous 4-component BVP for
even-parity polar metric and relativistic EC matter perturbations, sum all 9
multipole mode pairs for $2\le\ell\le4$, evolve the $(1+1)$D Eddington--Finkelstein
wave PDE to extract time-domain waveforms at $r_{\rm ext}=50M_0,80M_0,120M_0$,
and simultaneously integrate the coupled $(r_{\rm orb}(v),\Phi_{\rm GW}(v))$
orbit-phase equations over $v\in[0,200M_0]$. At $M_0\Omega=8^{-3/2}$, the
$(2,\pm2)$ pair supplies $84.54\%$ of the 9-mode total flux
$\mathcal F_E^{\rm tot}/(\mu/M_0)^2=1.9496\times10^{-4}$, the paired EC versus
zero-spin $(2,2)$ shift is $-1.4945\times10^{-8}$ (with boundary drift
$9.05\times10^{-15}$), the source flux shift reaches $+3.013\times10^{-5}$ at
$v=200M_0$ (yielding $+1.557\times10^{-5}$ extracted at $r_{\rm ext}=50M_0$ at
$v_{\rm obs}=200M_0$ and $+2.956\times10^{-5}$ at $r_{\rm ext}=120M_0$ at
retarded arrival $v_{\rm obs}=435.92M_0$), and the coupled time-dependent phase
shift over $v\in[0,200M_0]$ is $+1.367\times10^{-3}\text{ rad}$.
\end{abstract}
\maketitle
\section{Introduction}
\label{sec:introduction}

Extreme-mass-ratio inspirals (EMRIs) provide one of the cleanest strong-field
laboratories for testing gravity and probing the environment of massive black
holes.  In a typical EMRI, a stellar-mass compact object of mass
\(\mu\sim 1-100M_\odot\) slowly inspirals into a massive black hole of mass
\(M\sim 10^5-10^7M_\odot\), producing a mass ratio
\begin{equation*}
        \eta=\frac{\mu}{M}\ll 1 .
\end{equation*}
Because the inspiral is slow, the compact object spends a very large number of
cycles in the strong-field region before plunge.  The accumulated
gravitational-wave phase is therefore highly sensitive to small perturbations
of the background geometry, the orbital potential, and the radiation-reaction
balance.  This makes EMRIs particularly powerful probes of near-horizon
physics, environmental effects, and possible deviations from general
relativity.  LISA and Taiji mission studies motivate space-based observations of these
systems \cite{LISAProposal2017,HuWu2017}. EMRI waveform and
parameter-estimation studies quantify their information content
\cite{BarackCutler2004,AmaroSeoane2007,BabakEtAl2017,KatzEtAl2021}.
Multipole mapping and non-Kerr waveform studies investigate tests of
compact-object geometry
\cite{Ryan1995,CollinsHughes2004,MuguruzaSopuerta2026Kerr,
MuguruzaSopuerta2026Horizon}.

In general relativity, the leading description of an EMRI is obtained by
treating the secondary as a small body moving in the Kerr spacetime of the
central massive black hole, with its long-term inspiral driven by gravitational
radiation reaction.  However, the precision of an EMRI waveform also implies
that any additional matter distribution, dark-sector interaction, or modified
gravitational coupling near the massive black hole may leave a secular imprint
on the waveform phase.  Even when the local correction to the metric is small,
the accumulated phase shift may become observable after integration over
\(10^4-10^5\) orbital cycles.  For this reason, EMRIs have been widely studied
as probes of black-hole multipole moments, scalar fields, dark-matter
distributions, accretion environments, modified gravity, and horizon-scale
structure.

A particularly important environmental effect is the possible formation of a
dark-matter spike around a massive black hole.  If a black hole grows
adiabatically inside a pre-existing dark-matter halo, the central density
profile may be steepened into a spike.  The original analysis by
Gondolo and Silk showed that such adiabatic growth can enhance the inner
dark-matter density and produce a power-law spike near the black hole
\cite{GondoloSilk1999}.  In Newtonian language, a typical spike profile may be
written as
\begin{equation*}
        \rho_\chi(r)
        =
        \rho_0
        \left(
        \frac{r}{r_0}
        \right)^{-\gamma_{\rm sp}},
\end{equation*}
where \(\rho_0\) is the normalization, \(r_0\) is a reference radius, and
\(\gamma_{\rm sp}\) is the spike slope.  Relativistic calculations address strong-field changes in the spike
profile and capture cutoff \cite{SadeghianWill2013,Speeney2022}.
The influence of matter gravity and dynamical friction on inspiral
waveforms has been studied separately
\cite{Eda2013,BarausseCardosoPani2014}.  For motivation one may consider the representative profile
\begin{equation*}
        \rho_\chi(r)\propto r^{-3/2},
\end{equation*}
as an illustrative centrally concentrated dark component; it does not fix the metric used below.
Stellar relaxation can generate an \(r^{-3/2}\) dark-matter crest,
which is distinct from a steeper spike produced by adiabatic black-hole
growth \cite{MerrittHarfstBertone2007}. Multi-mass nuclear-cluster
relaxation and cumulative strong interactions with EMRIs can further
deplete the inner distribution \cite{SharpeEtAl2026DMSpikeDepletion}.
These long-term processes should not be identified with evolution over a
single observed inspiral.

Dark-matter spikes are usually modeled through their mass density, with effects
on inspirals from gravity, dynamical friction, and accretion. Spin-carrying
matter supplies a further theoretical motivation: in Einstein--Cartan gravity
intrinsic spin sources algebraic torsion
\cite{Hehl1976,Shapiro2002,Trautman2006,ObukhovKorotky1987}.
Eliminating torsion introduces local spin-dependent contributions to the
effective stress tensor. A quadratic spin density with an \(r^{-3}\) profile,
however, does not establish an \(r^{-3}\) metric response. Nor does one
component of that stress tensor determine a rotating geometry.

We quantify circular-orbit changes and conditional phase responses for a
specified near-zone Kerr deformation. The equatorial metric, exterior
domain, angular completion, and radiation prescriptions are stated
explicitly. The coefficient \(\alpha\) is an independent effective
parameter. The spherical Einstein--Cartan calculations in
Appendices~\ref{app:ec_source}, \ref{app:ec_accretion}, and
\ref{app:finite_cloud} establish the coarse-grained six-state Kramers-bundle spin
closure, static equilibrium obstructions, transonic inflow, and the nonlinear
finite-cloud evolution with its coupled dominant $(\ell,m)=(2,2)$ polar mode,
matter perturbations, and time-dependent total EMRI energy flux; they represent
the spherical dynamical sector rather than a direct high-spin ($a/M=0.9$)
derivation of the rotating metric. Table~\ref{tab:calculation_scope}
summarizes the quantities each calculation establishes.

The distinction between geometry and radiation is equally essential.
Relativistic nonvacuum perturbation calculations couple metric and matter
perturbations, and their radiation cannot generally be recovered by inserting
a modified orbital frequency into a vacuum quadrupole expression
\cite{Cardoso2022,Speeney2024}. Multi-parameter and post-adiabatic treatments
make this requirement explicit \cite{DattaMaselli2026,RahmanTakahashi2026}.
In the rotating benchmark we use a quadrupole closure for illustrative
trajectories and independently evaluate a fixed-frequency conservative
diagnostic with numerical Kerr flux data, while in the spherical finite-cloud
EC sector (Appendix~\ref{app:finite_cloud}) we directly solve the coupled
polar $(\ell,m)=(2,2)$ and higher-multipole gravitational and fluid
perturbation equations to obtain the time-dependent total EMRI energy flux.

\begin{figure*}[t]
\centering\includegraphics[width=.90\textwidth]{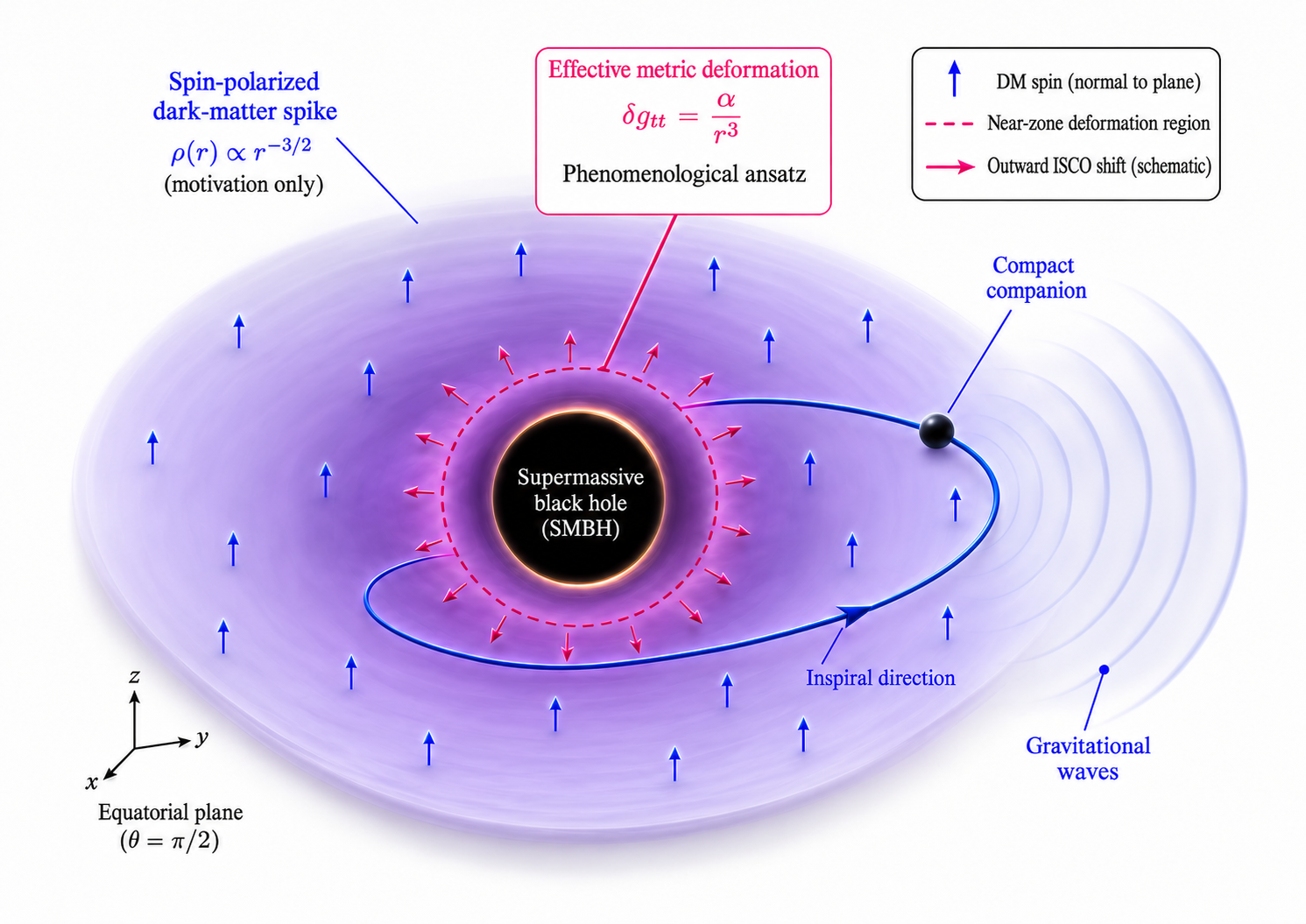}
\caption{Schematic motivation for the near-zone model. A compact companion
orbits a massive central object within an illustrative polarized dark-matter
spike and emits gravitational waves. The blue spin arrows are normal to the equatorial plane, consistent with
the local tensor orientation in Eq.~\eqref{eq:spin_components}. The density
and polarization remain motivation only; no source-to-metric map is assumed. Magenta arrows denote
the benchmark's outward ISCO shift schematically, not an outward force or
a derived spin--torsion barrier. The false-color equatorial density overlay,
central light ring, and wavefronts are illustrative, not observations or
ray-tracing results for the prescribed metric.}
\label{fig:scope}
\end{figure*}

Section~\ref{sec:RC_spin_fluid} introduces the spin-fluid motivation and states
its limitations. Section~\ref{sec:metric_perturbation_spin_spike} distinguishes
source response from the adopted metric and specifies the exterior completion.
Section~\ref{sec:orbits} develops the orbital and flux diagnostics.
Section~\ref{sec:dephasing} reports recomputed trajectories, phases, and
waveform illustrations. Section~\ref{sec:observations} discusses the conditions
needed for an observational interpretation, and Sec.~\ref{sec:conclusions}
summarizes the results.

\begin{table*}[t]
\caption{Scope of the calculations. The fluid examples do not
identify their statistical spin coefficient with the rotating metric's
independent deformation parameter.}
\label{tab:calculation_scope}
\begin{ruledtabular}
\begin{tabular}{p{.18\textwidth}p{.34\textwidth}p{.40\textwidth}}
Calculation & Background and matter assumptions & Established quantities\\
\hline
Rotating benchmark & Prescribed exterior metric, Eq.~\eqref{eq:full_metric};
metric-coupled, unspinning secondary & Circular geodesics, stability checks,
and phase diagnostics under specified flux prescriptions. No derived EC
source or deformed-background radiation flux.\\[3pt]
Static source test & Averaged, unpolarized, isotropic spin fluid;
explicit barotropic closure & Local TOV equilibria and conditional
boundary/compressibility obstructions. No completed black-hole atmosphere.\\[3pt]
Inflow source test & The same functional fluid family with separate
diagnostic parameters; transonic inflow from a finite reservoir &
Leading Schwarzschild matter flow and its first metric backreaction.
The computed inflow is matter energy, not EMRI gravitational-wave power.\\[3pt]
Finite-cloud \& coupled $(2,2)$+matter EMRI evolution & Six-state
coarse-grained Kramers-bundle EC closure; nonlinear finite cloud with vacuum
exterior; simultaneous 4-component polar+matter BVP and $(1+1)$D wave PDE &
Direct 4D tensor contraction $\tfrac12 s_{\mu\nu}s^{\mu\nu}=\chi_s n^2$,
simultaneous 4-component $(\ell,m)=(2,2)$+matter BVP, complete 9-mode
($2\le\ell\le4$) sum, $(1+1)$D PDE waveform extraction at $r_{\rm ext}=50,80,120M_0$,
and simultaneous coupled $(r_{\rm orb}(v),\Phi_{\rm GW}(v))$ evolution.
\end{tabular}
\end{ruledtabular}
\end{table*}

\section{Riemann--Cartan Geometry and Weyssenhoff Spin Fluid}
\label{sec:RC_spin_fluid}
\subsection{Geometrical and matter conventions}
The independent variables of minimal Einstein--Cartan theory are a metric and
a metric-compatible connection. With signature \((-+++)\) we write
\begin{equation}
\Gamma^\lambda{}_{\mu\nu}
=\mathring\Gamma^\lambda{}_{\mu\nu}+K^\lambda{}_{\mu\nu},\qquad
Q^\lambda{}_{\mu\nu}=\Gamma^\lambda{}_{[\mu\nu]},
\label{eq:connection}
\end{equation}
where \(K\) is contortion and the alternative torsion convention is
\(T^\lambda{}_{\mu\nu}=2Q^\lambda{}_{\mu\nu}\). The action has the form
\begin{equation}
S=\frac{1}{2\kappa}\int d^4x\sqrt{-g}\,R(g,\Gamma)
+S_{\rm m}[g,\Gamma,\Psi],\quad\kappa=8\pi G.
\label{eq:EC_action}
\end{equation}
The connection equation is algebraic in torsion for this minimal action.
Torsion vanishes in spinless vacuum; there is no additional independent
vacuum torsion wave in this theory \cite{Hehl1976,Shapiro2002,Trautman2006}.
These standard facts motivate the discussion, rather than provide a field
equation for the numerical metric. For Dirac matter, elimination of torsion
produces a local nonlinear spinor interaction \cite{HehlDatta1971};
in Appendix~\ref{app:finite_cloud}, we construct an explicit
six-state Kramers-bundle spin representation on a Fermi--Walker-transported
comoving tetrad that satisfies four-dimensional MPD spin transport and direct
coordinate contraction $\tfrac12 s_{\mu\nu}s^{\mu\nu}=\chi_s n^2$, showing
why a single polarized branch breaks spherical isotropy while the six-state
coarse-grained ensemble closes the isotropic fluid with
$\langle\mathcal S^2\rangle=\chi_s n^2$ and $\delta\langle\mathcal S^2\rangle=2\chi_s n\delta n$.

For a Weyssenhoff fluid one introduces
\begin{equation}
S^\lambda{}_{\mu\nu}=u^\lambda s_{\mu\nu},\qquad
s_{\mu\nu}=-s_{\nu\mu},\qquad s_{\mu\nu}u^\nu=0.
\label{eq:frenkel}
\end{equation}
The last relation is the Frenkel constraint. The fluid action, spin evolution,
and conservation equations are additional ingredients, not consequences of
this algebraic constraint alone. We refer directly to the variational
treatment of Obukhov and Korotky \cite{ObukhovKorotky1987} for those equations.

\subsection{Local spin configuration and the source limitation}
A local kinematic spin configuration can be expressed by its nonzero component. In a Schwarzschild patch with \(f=1-2M/r\), take
\begin{equation}
u^\mu=(f^{-1/2},0,0,0),\qquad
s_{r\phi}=-s_{\phi r}=\sigma(r)\frac{r\sin\theta}{\sqrt f},
\label{eq:spin_components}
\end{equation}
all other independent spin components being zero. It follows that
\begin{equation}
s_{\mu\nu}u^\nu=0,\qquad
\mathcal S^2\equiv\tfrac12s_{\mu\nu}s^{\mu\nu}=\sigma^2.
\label{eq:spin_norm}
\end{equation}
These are kinematic checks only. This local orientation is not a demonstration
of spherical symmetry, regularity of a polarized field on the entire sphere,
or a solution of the spin-fluid equations. A statistically isotropic spin
ensemble and a fluid with net polarization are different matter models.

After eliminating torsion the metric equation can be organized as
\begin{equation}
\mathring G_{\mu\nu}=\kappa T^{\rm eff}_{\mu\nu},\qquad
T^{\rm eff}_{\mu\nu}=T^{\rm mat}_{\mu\nu}+U^{\rm spin}_{\mu\nu}.
\label{eq:effective_source}
\end{equation}
The split depends on the matter action and stress-tensor convention; spin
derivatives can occur as well as quadratic terms
\cite{ObukhovKorotky1987,Hehl1976}. Equations~\eqref{eq:frenkel} and
\eqref{eq:spin_norm} do not determine the complete right-hand side.
For illustration, the parameterization
\begin{equation}
U^{\rm spin}_{\hat0\hat0}=-\lambda_s\sigma^2,\quad
U^{\rm spin}_{tt}=-f\lambda_s\sigma^2,\quad
U^t{}_{t}=+\lambda_s\sigma^2
\label{eq:source_indices}
\end{equation}
is only a conditional illustrative component, with an unspecified coefficient
\(\lambda_s>0\). It is not a derived Weyssenhoff stress tensor. Neither its
normalization nor a force interpretation is inferred from the Frenkel
condition. We do not use it as a source for the adopted Kerr deformation.

A complete source-based construction must specify the fluid action, pressures,
microscopic spin states, and boundary conditions, solve all components of
Eq.~\eqref{eq:effective_source}, and verify
\(\mathring\nabla_\mu T^{\mu\nu}_{\rm eff}=0\) together with pointwise
MPD spin transport. In the spherical dynamic sector,
Appendix~\ref{app:finite_cloud} completes this microscopic construction and
evolves the coupled metric, matter, and $(\ell,m)=(2,2)$ radiation system,
whereas in the high-spin rotating sector ($a/M=0.9$) the main text treats the
near-zone metric as a phenomenological input.

\section{Spin Source, Linearized Response, and Near-Zone Deformation}
\label{sec:metric_perturbation_spin_spike}
\subsection{Density motivation and the corrected static response}
A dark-matter spike is often parameterized as
\begin{equation}
\rho_\chi(r)=\rho_0(r/r_0)^{-\gamma_{\rm sp}},\qquad
\sigma(r)=\sigma_0 r^{-3/2},
\label{eq:spin_profile}
\end{equation}
where the second relation is an illustrative spin profile, not a solved fluid
equilibrium. A microscopic polarized fermion model would have
\(s\simeq\hbar\mathcal P\rho_\chi/(2m_\chi)\), with \(0\leq\mathcal P\leq1\).
Here \(\mathcal P\) denotes a net polarization fraction; it is distinct
from the statistical EOS coefficient \(\zeta\) in the unpolarized
appendices. Both the density and polarization require independent
justification.

The difference between a source and a metric tail can be seen without claiming
a physical source solution. For a one-function static ansatz,
\begin{equation}
ds^2=-Fdt^2+F^{-1}dr^2+r^2d\Omega_2^2,\quad
F=1-\frac{2M}{r}+\varepsilon h,
\label{eq:static_ansatz}
\end{equation}
direct evaluation of the mixed Einstein tensor gives
\begin{equation}
G^t{}_t=G^r{}_r=\frac{rF'+F-1}{r^2},\qquad
\delta G^t{}_t=\frac{(rh)'}{r^2}.
\label{eq:mixed_einstein}
\end{equation}
Here \(\delta G^t{}_t\) is the coefficient of \(\varepsilon\),
namely \(\partial_\varepsilon G^t{}_t|_{\varepsilon=0}\).
Using the mixed component in Eq.~\eqref{eq:source_indices} and
\(A=\kappa\lambda_s\sigma_0^2\), the conditional equation and its integral are
\begin{equation}
\frac{(rh)'}{r^2}=\frac{A}{r^3},\qquad
h=\frac{C_m}{r}+\frac{A\ln(r/r_m)}{r}.
\label{eq:correct_static}
\end{equation}
The integration constant \(C_m\) is distinct from the spin-fluid
coefficient \(C\) in Appendix~\ref{app:ec_source}.
Equation~\eqref{eq:correct_static} remains conditional: the one-function
ansatz additionally requires \(T^t{}_t=T^r{}_r\), which has not been
derived from a Weyssenhoff model. A cutoff and outer matching would also be
needed to interpret its logarithmic tail globally. Matching an amplitude or
force at one radius does not determine a rotating geometry. This example
only demonstrates why a source's radial power law cannot be identified with
the same power law in the metric.

For a general static anisotropic effective source the appropriate independent
functions obey, in \(G=c=1\),
\begin{align}
ds^2&=-e^{2\nu}dt^2+\frac{dr^2}{1-2m/r}+r^2d\Omega_2^2,\nonumber\\
m'&=4\pi r^2\rho_{\rm eff},\qquad
\nu'=\frac{m+4\pi r^3p_r}{r(r-2m)},\nonumber\\
p_r'&=-(\rho_{\rm eff}+p_r)\nu'+\frac{2(p_t-p_r)}r.
\label{eq:full_static_requirements}
\end{align}
Appendix~\ref{app:ec_source} closes and solves a separate spherical,
statistically unpolarized spin-fluid model locally, and tests its inner and
outer completion. The ordinary positive-pressure branch cannot form an
unsupported hollow atmosphere with two vacuum boundaries; the illustrative
nonzero-spin polytropic branches encounter a finite-density compressibility
endpoint before any demonstrated horizon. These results do not supply an
EC source for the rotating metric introduced next.
Appendix~\ref{app:ec_accretion} instead allows radial inflow and constructs
a horizon-regular transonic source with its first dynamical metric
backreaction, within an explicit finite-reservoir model.
Appendix~\ref{app:finite_cloud} then replaces continued reservoir supply by a
finite cloud matched to an exterior vacuum, establishes the six-state
coarse-grained Kramers-bundle MPD spin closure, and solves the simultaneous 4-component
time-dependent $(\ell,m)=(2,2)$ polar, axial, and fluid perturbation equations
for the total EMRI energy flux.

\subsection{Four-dimensional definition and dimensions}
Define a stationary,
axisymmetric exterior metric in Boyer--Lindquist coordinates by
\begin{align}
\Sigma&=r^2+a^2\cos^2\theta,\quad \Delta=r^2-2Mr+a^2,\nonumber\\
g_{tt}&=-1+\frac{2Mr}{\Sigma}+\frac{\alpha}{r^3},\nonumber\\
g_{t\phi}&=-\frac{2Mar\sin^2\theta}{\Sigma}
-\frac{\alpha aM}{r^4}\sin^2\theta,\nonumber\\
g_{\phi\phi}&=\left(r^2+a^2+\frac{2Ma^2r\sin^2\theta}{\Sigma}\right)
\sin^2\theta+\frac{\alpha M^2}{r^3}\sin^4\theta,\nonumber\\
g_{rr}&=\frac{\Sigma}{\Delta},\qquad g_{\theta\theta}=\Sigma.
\label{eq:full_metric}
\end{align}
All other components vanish except for metric symmetry. Here \([a]=[M]=L\)
and \([\alpha]=L^3\). Thus the added \(tt,t\phi,\phi\phi\) components
have dimensions \(1,L,L^2\), respectively. The explicit factors of \(M\)
ensure the correct dimensions when restoring the \(M=1\) notation.

The angular factors are an explicit completion chosen for this calculation;
they are not unique consequences of an action. They preserve the equatorial
components exactly and give reflection symmetry about \(\theta=\pi/2\).
The spin-independent angular deformation is allowed as a phenomenological
choice; even at \(a=0\) this full four-dimensional metric need not be
spherically symmetric. No Carter constant, generic-orbit separability, or
rotating field-equation solution is assumed.

Below we set \(M=1\), so that the symbols \(r,t,a,\alpha\) denote
\(r/M,t/M,a/M,\alpha/M^3\). In particular, on the equator
\begin{align}
g_{tt}&=-1+2/r+\alpha/r^3,\nonumber\\
g_{t\phi}&=-2a/r-\alpha a/r^4,\nonumber\\
g_{\phi\phi}&=r^2+a^2+2a^2/r+\alpha/r^3,\quad
g_{rr}=r^2/\Delta.
\label{eq:equatorial_metric}
\end{align}
This defines a specified
metric family. We evaluate this family directly; the small-\(\alpha\)
expansion is used for diagnostics rather than silently dropping all nonlinear
terms generated when computing geodesics.

\subsection{Exterior domain, signature, and force sign}
The demonstrated domain is \(a=0.9\), \(|\alpha|\leq10^{-3}\), \(r\geq2\).
It covers every orbit used here and stays outside the Kerr coordinate horizon
\(r_+=1+\sqrt{1-a^2}\). A horizon-penetrating extension is not supplied.
In particular, retaining Kerr \(g_{rr}\) does not prove that its coordinate
pole is a regular horizon of the deformed metric.

On this domain \(\Sigma,\Delta>0\). At the axis,
\begin{equation}
\lim_{\theta\to0,\pi}
\frac{g_{\phi\phi}}{g_{\theta\theta}\sin^2\theta}=1,
\qquad g_{t\phi}=O(\sin^2\theta),
\label{eq:axis}
\end{equation}
so the angular addition creates no conical defect for period \(2\pi\).
The remaining signature test is
\(D\equiv g_{t\phi}^2-g_{tt}g_{\phi\phi}>0\), with
\(g_{\phi\phi}>0\) away from the axis. For example, writing
\(B_K=g^{K}_{t\phi}/\sin^2\theta\) and
\(C_K=g^{K}_{\phi\phi}/\sin^2\theta\), the bounds
\(|B_K|\leq2a/r\), \(|g^K_{tt}|\leq1\), and
\(C_K\leq r^2+a^2+2a^2/r\) imply
\begin{equation}
\left|\frac{D}{\sin^2\theta}-\Delta\right|<9.29\times10^{-4}.
\label{eq:signature_bound}
\end{equation}
Since \(\Delta\geq0.81\), the signature is Lorentzian throughout this domain.
Also \(g_{\phi\phi}/\sin^2\theta>4.8098\), so the azimuthal circles
are spacelike. A \(500\times301\) grid over \(2\leq r\leq100\)
at \(\alpha=-10^{-3},0,10^{-3}\) independently gives
\(\min D/\sin^2\theta=0.809399\). Smooth coefficients and a nonzero
determinant in regular axis charts establish local exterior regularity only.

The weak-field potential must use the conventional sign:
\begin{equation}
g_{tt}\simeq-(1+2\Phi_N),\qquad
\Phi_N=-\frac Mr-\frac{\alpha}{2r^3}
\label{eq:potential_sign}
\end{equation}
for the nonrotating \(tt\) sector with dimensions restored. Consequently
\(\alpha>0\) contributes an inward Newtonian radial acceleration
\(-3\alpha/(2r^4)\). An outward ISCO displacement is not a proof of a
repulsive force. The positive benchmark is used without an outward-force interpretation.

The exterior-domain statement is essential. Directly extending the same
formula inward at the benchmark gives \(D=0\) on the equator at
\(r\simeq1.43698M\), outside the Kerr coordinate value
\(r_+\simeq1.43589M\). This matrix degeneracy lies outside the stated
\(r\geq2M\) domain; its geometric interpretation would require a separate
extension and curvature analysis. We do not impose a horizon boundary
condition on the deformed geometry or calculate its horizon flux.

The angular completion also contains physical freedom. For example,
\begin{equation}
\delta g_{tt}=\beta\frac{\alpha}{r^3}\cos^2\theta
\label{eq:angular_freedom}
\end{equation}
in dimensionless units leaves the equatorial metric and its radial
derivatives unchanged, but changes the vertical frequency by
\(\delta\Omega_\theta^2=-\beta\alpha/r^5\). Thus all equatorial
results below are independent of this particular freedom, whereas the
vertical-stability sample applies only to the completion actually specified.

\subsection{Comparison with a parameterized Kerr geometry}
To test whether the leading \(tt\) tail characterizes the orbital response,
we compare with the \(\epsilon_3\)-only Johannsen metric
\cite{Johannsen2013}. Setting its other deviation functions to their Kerr
values gives the conformal form
\begin{equation}
g^{J}_{\mu\nu}=\left(1+\frac{\epsilon_3M^3}{r\Sigma}\right)
g^{K}_{\mu\nu}.
\label{eq:johannsen_comparator}
\end{equation}
On the equator, choosing \(\epsilon_3=-\alpha/M^3\) matches the leading
asymptotic correction \(\delta g_{tt}=+\alpha/r^3+O(r^{-4})\).
The remaining components differ, so this is a comparison prescription, not
an identification or a uniqueness claim. Both families have the same
asymptotic mass, spin, and time normalization. The energy-frequency relation
is consequently a more informative comparison than a coordinate radius
alone. The calculation below tests the specified full equatorial deformation
against this known family; it does not claim a new general parameterization.

\section{Orbital Dynamics and Effective Deformation}
\label{sec:orbits}
We use circular equatorial geodesics of metric~\eqref{eq:equatorial_metric}. All formulae in this and the following
section use the dimensionless \(M=1\) notation just defined. The secondary
has no spin, has fixed mass, and couples to the metric through
\(S_p=-\mu\int ds\). Its leading motion is therefore a Levi--Civita
geodesic of this prescribed metric. There is no independent torsion connection
specified for the numerical problem.
\subsection{Shift of the ISCO}
\label{subsec:isco_shift}

The conserved specific energy and angular momentum of a test body moving on
the equatorial plane are
\begin{equation}
        E
        =
        -u_t
        =
        -g_{tt}u^t-g_{t\phi}u^\phi,
        \qquad
        L_z
        =
        u_\phi
        =
        g_{t\phi}u^t+g_{\phi\phi}u^\phi .
\end{equation}
Solving these equations for \(u^t\) and \(u^\phi\) gives
\begin{align}
        u^t
        &=
        \frac{
        Eg_{\phi\phi}+L_zg_{t\phi}
        }{
        g_{t\phi}^2-g_{tt}g_{\phi\phi}
        },
        \\
        u^\phi
        &=
        -
        \frac{
        Eg_{t\phi}+L_zg_{tt}
        }{
        g_{t\phi}^2-g_{tt}g_{\phi\phi}
        } .
\end{align}
Using the normalization \(u^\mu u_\mu=-1\), the radial equation can be written
as
\begin{equation}
        g_{rr}\left(\frac{dr}{d\tau}\right)^2
        =
        V_r(r;E,L_z,a,\alpha),
\end{equation}
where
\begin{equation}
        V_r(r;E,L_z,a,\alpha)
        =
        \frac{
        E^2g_{\phi\phi}
        +
        2EL_zg_{t\phi}
        +
        L_z^2g_{tt}
        }{
        g_{t\phi}^2-g_{tt}g_{\phi\phi}
        }
        -1 .
\end{equation}
Here \(\tau\) is proper time; balance-law time derivatives below refer to
coordinate time \(t\). This quantity plays the role of the radial effective potential.  Circular
orbits satisfy
\begin{equation}
        V_r=0,
        \qquad
        \frac{\partial V_r}{\partial r}=0,
\end{equation}
and the ISCO is obtained from the additional marginal-stability condition
\begin{equation}
        \frac{\partial^2 V_r}{\partial r^2}=0 .
\end{equation}
Equivalently, for the circular sequence one may locate the ISCO by the
minimum of the specific energy \(E(r)\),
\begin{equation}
        \left.
        \frac{dE(r,a,\alpha)}{dr}
        \right|_{r=r_{\rm ISCO}}
        =
        0 .
\end{equation}

The angular frequency of a prograde circular orbit follows directly from the
stationary-axisymmetric circular-orbit condition,
\begin{equation}
        \Omega_\phi
        =
        \frac{
        -\partial_r g_{t\phi}
        +
        \sqrt{
        \left(\partial_r g_{t\phi}\right)^2
        -
        \left(\partial_r g_{tt}\right)
        \left(\partial_r g_{\phi\phi}\right)}
        }{
        \partial_r g_{\phi\phi}
        }.
\end{equation}
For \(\alpha=0\), this reduces to the standard Kerr result
\begin{equation}
        \Omega_{\rm K}
        =
        \frac{1}{r^{3/2}+a}.
\end{equation}
The corresponding energy and angular momentum are
\begin{align}
        E(r,a,\alpha)
        &=
        -
        \frac{
        g_{tt}+g_{t\phi}\Omega_\phi
        }{
        \sqrt{
        -g_{tt}
        -
        2g_{t\phi}\Omega_\phi
        -
        g_{\phi\phi}\Omega_\phi^2}
        },
        \\
        L_z(r,a,\alpha)
        &=
        \frac{
        g_{t\phi}
        +
        g_{\phi\phi}\Omega_\phi
        }{
        \sqrt{
        -g_{tt}
        -
        2g_{t\phi}\Omega_\phi
        -
        g_{\phi\phi}\Omega_\phi^2}
        } .
\end{align}
Expanding in \(\alpha\), one writes
\begin{align}
        \Omega_\phi(r,a,\alpha)
        &=
        \Omega_{\rm K}(r,a)
        +
        \alpha\Omega_1(r,a)
        +
        O(\alpha^2),
        \\
        E(r,a,\alpha)
        &=
        E_{\rm K}(r,a)
        +
        \alpha E_1(r,a)
        +
        O(\alpha^2),
        \\
        L_z(r,a,\alpha)
        &=
        L_{z{\rm K}}(r,a)
        +
        \alpha L_{z1}(r,a)
        +
        O(\alpha^2).
\end{align}
The Kerr terms are the standard prograde equatorial expressions
\cite{BardeenPressTeukolsky1972},
\begin{equation}
        E_{\rm K}
        =
        \frac{
        r^{3/2}-2r^{1/2}+a
        }{
        r^{3/4}
        \sqrt{
        r^{3/2}-3r^{1/2}+2a
        }
        },
\end{equation}
and
\begin{equation}
        L_{z{\rm K}}
        =
        \frac{
        r^2-2ar^{1/2}+a^2
        }{
        r^{3/4}
        \sqrt{
        r^{3/2}-3r^{1/2}+2a
        }
        } .
\end{equation}

For \(a=0.9\) and \(\alpha=10^{-3}\), direct evaluation gives
\begin{align}
r_{\rm ISCO}^{K}&=2.320883042,\nonumber\\
r_{\rm ISCO}^{\alpha}&=2.324536881,\nonumber\\
\Delta r_{\rm ISCO}&=3.6538391\times10^{-3}.
\label{eq:isco_numeric}
\end{align}
The shift is a property of the specified equatorial metric. Evaluating
\(V_r,\partial_rV_r,\partial_r^2V_r\), with \(E,L_z\) held fixed in the
partial derivatives, gives residuals below \(4\times10^{-15}\) at both
marginal orbits. These derivatives check radial stability. Equatorial
reflection symmetry ensures the equator is a geodesic submanifold; vertical
stability is checked separately in Appendix~\ref{app:verification}.

\begin{figure}[tb]
\centering\includegraphics[width=\linewidth]{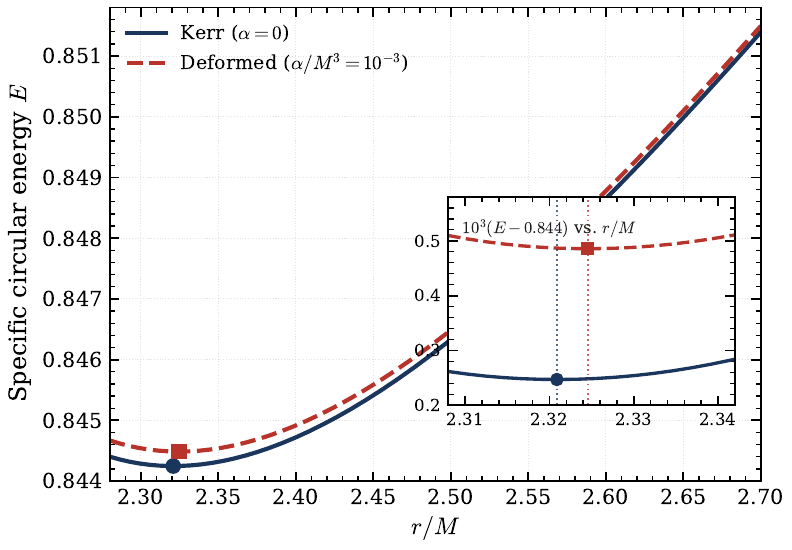}
\caption{Specific energy along circular equatorial sequences. Dots locate the radially marginal orbits.
The small outward shift is a property of the metric, not an identified
spin-fluid force.}\label{fig:energy}
\end{figure}

Coordinate ISCO radii depend on the radial gauge. For the benchmark we also
find
\begin{align}
M\Omega_{\rm ISCO}^{K}&=0.225441709,\nonumber\\
M\Omega_{\rm ISCO}^{\alpha}&=0.225050024,\nonumber\\
E_{\rm ISCO}^{K}&=0.844247008,\qquad
E_{\rm ISCO}^{\alpha}=0.844485526.
\label{eq:isco_invariants}
\end{align}
These quantities use the asymptotically normalized time and specific energy.
The spin dependence is shown in Table~\ref{tab:spin_scan}; both signs of
the small deformation are evaluated in the companion calculation.

\begin{table}[tb]
\caption{Circular marginal-orbit changes for \(\alpha/M^3=10^{-3}\),
relative to Kerr at the same central spin and mass. The background checks
at \(a/M=0.9\) do not constitute a global-domain proof for other spins.}
\label{tab:spin_scan}
\begin{ruledtabular}\begin{tabular}{crrr}
\(a/M\)&\(\Delta r/M\)&\(M\Delta\Omega\)&\(\Delta E\)\\\hline
0 & \(5.3671\times10^{-4}\)&\(-7.7050\times10^{-6}\)&\(4.4146\times10^{-6}\)\\
0.5 & \(1.0237\times10^{-3}\)&\(-3.2548\times10^{-5}\)&\(1.7437\times10^{-5}\)\\
0.9 & \(3.6538\times10^{-3}\)&\(-3.9168\times10^{-4}\)&\(2.3852\times10^{-4}\)\\
\end{tabular}\end{ruledtabular}
\end{table}

\subsection{Secondary spin and self-force scope}
The spinless assumption defines the calculation; it does not estimate away
spin effects. A spinning secondary would require a specified worldline action,
spin supplementary condition, and multipolar expansion. Standard
Mathisson--Papapetrou--Dixon theory illustrates those additional ingredients
\cite{Mathisson1937,Papapetrou1951,Dixon1970,Tulczyjew1959}; it does not
supply spin--torsion couplings for an unspecified matter theory. A general Riemann--Cartan spinning-body evolution is therefore outside the
model defined here.

Writing \(\eta=\mu/M\) and \(\epsilon=\alpha/M^3\) before setting \(M=1\),
a source-based two-parameter expansion would have the structure
\begin{equation}
g=g^K+\epsilon h^{\rm env}+\eta h^{(1)}
+\epsilon\eta h^{\rm mix}+\cdots.
\label{eq:two_parameter}
\end{equation}
For a consistent adiabatic calculation the phase scales schematically as
\begin{equation}
\Phi=\eta^{-1}\Phi_0(\epsilon)+\Phi_1(\epsilon)+O(\eta).
\label{eq:phase_order}
\end{equation}
The adiabatic and post-adiabatic ordering follows the two-timescale
organization of EMRI motion \cite{HindererFlanagan2008}.
The mixed perturbation can change the leading flux and hence the phase at
\(O(\epsilon/\eta)\). It cannot generally be discarded as a higher-order
phase effect \cite{DattaMaselli2026,RahmanTakahashi2026}. In the rotating
phenomenological benchmark we bracket this dissipative sector with explicit
flux templates, whereas in the spherical finite-cloud EC sector
(Appendix~\ref{app:finite_cloud}) we solve the coupled first-order in $\eta$,
nonlinear in background density metric--matter perturbation system explicitly.
Second-order in $\eta$ self-force contributions \cite{BarackPound2019,PoundWardell2021}
remain beyond the leading adiabatic and time-dependent balance equations.

\subsection{Energy and angular-momentum fluxes}
The balance equation is
\begin{equation}
\frac{dr}{dt}=-\frac{\mathcal F_E}{E_{,r}},\qquad
\mathcal F_E=-\frac{dE}{dt}>0.
\label{eq:balance}
\end{equation}
Here \(E\) is specific orbital energy; a physical power must be divided by
the secondary mass when converting to the same specific-energy normalization
and time units. Thus the leading flux carries one factor of \(\eta\),
not \(\eta^2\), in Eq.~\eqref{eq:balance} with dimensionless time.
For a specified physical theory one would consistently account for
\begin{equation}
\mathcal F_E=\mathcal F_\infty+\mathcal F_H
+\mathcal F_{\rm matter}+\mathcal F_{\rm additional}.
\label{eq:flux_channels}
\end{equation}
Terms denote energy reaching infinity, the central boundary, matter, or other
degrees of freedom; they need not all be nonzero or positive. Energy exchange
already included in a perturbative balance must not be counted again as a
separate drag term. For the rotating phenomenological metric in the main text,
these channels are represented by explicit baseline and template prescriptions;
for the self-gravitating spherical EC cloud in Appendix~\ref{app:finite_cloud},
all three physical channels---$\mathcal F_\infty$, $\mathcal F_H$, and the
coupled relativistic acoustic matter flux $\mathcal F_{\rm matter}$---are
solved explicitly across the dominant $(\ell,m)=(2,2)$ mode and higher
multipoles ($\ell\le4$).

For illustrative trajectories we use the explicit closure
\begin{equation}
\mathcal F_Q(r;\alpha)=\frac{32}{5}\eta r^4\Omega(r;\alpha)^6.
\label{eq:quadrupole}
\end{equation}
It has the weak-field circular quadrupole normalization
\cite{PetersMathews1963}. It is not a calculated nonvacuum flux, and is
not an accurate strong-field Kerr flux even at \(\alpha=0\). Matching the
analytic Kerr geodesics below validates the conservative limit only.
On a circular sequence \(dE=\Omega\,dL_z\); an angular-momentum loss
consistent with this closure is \(\mathcal F_L=\mathcal F_Q/\Omega\).

For the conservative diagnostic we freeze the radiation kernel at a shared
frequency. Write \(r_K(\Omega)=(\Omega^{-1}-a)^{2/3}\) and distinguish
\begin{align}
\mathcal F_{Q0}(\Omega)&=\frac{32}{5}\eta r_K(\Omega)^4\Omega^6,\nonumber\\
\mathcal F_K(\Omega)&=\eta\left[f_\infty(r_K)+f_H(r_K)\right].
\label{eq:kerr_flux_baseline}
\end{align}
Kerr radiation is computed through black-hole perturbation theory,
whose separated field equations were developed by Teukolsky
\cite{Teukolsky1973}. The numerical Kerr kernels \(f_\infty,f_H\)
used here are the circular equatorial data provided by S.~A.~Hughes for Ref.~\cite{Taracchini2014} and distributed
by the Black Hole Perturbation Toolkit \cite{BHPToolkitData}. Their
normalization is the physical power divided by \(\mu^2/M^2\); multiplying
by \(\eta\) gives the specific-energy loss per dimensionless time.
The horizon contribution is negative over the interval used, as expected
for superradiant extraction, and is already included once in
\(\mathcal F_K\). We interpolate the positive total kernel, not its two
signed components separately.

The numerical baseline is held identical in both geometries at the same
\(\Omega\). This defines \(\mathcal F_1=0\) as a conservative diagnostic;
it does not transfer the physical Kerr horizon or perturbation equations to
the deformed metric. To expose unknown radiation dependence, set
\begin{equation}
\mathcal F(\Omega;\alpha,b)=\mathcal F_0(\Omega)
\left[1+\alpha b(\Omega/\Omega_f)^2\right],
\label{eq:flux_template}
\end{equation}
where \(\mathcal F_0\) is either baseline in
Eq.~\eqref{eq:kerr_flux_baseline}. Values \(b=\pm1\) change the
flux by at most \(10^{-3}\) at the benchmark and are sensitivity
scenarios, not predictions or uncertainty bounds. Table~\ref{tab:fixed_frequency}
uses both baselines explicitly.

\subsection{Fixed-frequency phase and conservative--dissipative split}
Since \(\dot\Omega=-\mathcal F_E/E_{,\Omega}\),
\begin{align}
T&=-\int_{\Omega_i}^{\Omega_f}\frac{E_{,\Omega}}{\mathcal F_E}\,d\Omega,
\nonumber\\
\Phi_{\rm GW}&=-2\int_{\Omega_i}^{\Omega_f}
\frac{\Omega E_{,\Omega}}{\mathcal F_E}\,d\Omega.
\label{eq:phase_frequency}
\end{align}
For each chosen flux prescription, define
\begin{align}
\Delta\Phi_\Omega&=\Phi_{\rm GW}^{\alpha}(\Omega_i,\Omega_f)
-\Phi_{\rm GW}^{K}(\Omega_i,\Omega_f),\nonumber\\
\Delta T&=T_\alpha(\Omega_i,\Omega_f)-T_K(\Omega_i,\Omega_f).
\label{eq:matched_phase_definition}
\end{align}
For \(E=E_0+\alpha E_1\), \(\mathcal F_E=\mathcal F_0+\alpha\mathcal F_1\)
at fixed endpoints, the leading change is
\begin{align}
\delta\Phi_{\rm GW}={}&-2\alpha\int_{\Omega_i}^{\Omega_f}\Omega
\left[\frac{E_{1,\Omega}}{\mathcal F_0}
-\frac{E_{0,\Omega}\mathcal F_1}{\mathcal F_0^2}\right]d\Omega\nonumber\\
&+O(\alpha^2).
\label{eq:phase_split}
\end{align}
The first and second terms define the conservative and dissipative diagnostic
contributions in this specified convention. Both scale as \(\alpha/\eta\).
Only the full phase of a consistent metric-and-radiation model can be promoted
to a physical waveform prediction; the split does not provide such a model.

\section{Gravitational Wave Emission and Dephasing}
\label{sec:dephasing}
We keep the fiducial mass and orbit choices
\begin{align}
M&=10^6M_\odot,\quad\eta=10^{-5},\nonumber\\
a&=0.9,\quad\alpha=10^{-3},\quad r_0=8.
\label{eq:fiducial}
\end{align}
Geodesics are evaluated directly from the prescribed metric. Derivatives are
analytic, and the balance integrals are evaluated along the circular sequence
instead of integrating a stiff \(dr/dt\) equation at its marginal endpoint.

Two matching conventions are used. The trajectory illustration starts
both models at the same radius \(r_0=8M\), sets their initial phases to
zero, and compares them at a shared coordinate time. The principal
fixed-frequency diagnostic instead matches \(\Omega_i\) and
\(\Omega_f\), allowing the radii and elapsed times to differ.
The former does not match the initial frequencies. These conventions
define different phase diagnostics and are not a detector-based alignment.

\subsection{Inspiral trajectory, adiabaticity, and termination}
For the quadrupole-closure illustration, both sequences start at \(r_0=8\)
with zero phase. Integration along a circular sequence gives
\begin{align}
t(r)&=\int_r^8\frac{E_{,r'}}{\mathcal F_Q}\,dr',\nonumber\\
\Phi(r)&=2\int_r^8\frac{\Omega E_{,r'}}{\mathcal F_Q}\,dr'.
\label{eq:radial_integrals}
\end{align}
This nonsingular quadrature representation does not remove the physical
failure of a circular adiabatic approximation as \(E_{,r}\to0\).
The transition-to-plunge regime must be distinguished from the adiabatic
inspiral \cite{OriThorne2000}.

Two local diagnostics are useful:
\begin{equation}
\mathcal A_r=\frac{2\pi|\dot r|}{r\Omega},\qquad
\mathcal A_\kappa=\frac{|\dot\kappa_r|}{\kappa_r^2},
\label{eq:adiabatic_diagnostics}
\end{equation}
where \(\kappa_r\) is the coordinate-time radial epicyclic frequency,
\begin{equation}
\kappa_r^2=-\frac{\partial_r^2V_r}{2g_{rr}(u^t)^2}
\label{eq:radial_epicycle}
\end{equation}
evaluated on a circular orbit, holding \(E,L_z\) fixed in the potential
derivative. Along the balance evolution,
\(\dot\kappa_r=\kappa_{r,r}\dot r\). Small values express a local
timescale hierarchy; they are not rigorous waveform-error bounds.

\begin{figure*}[t]
\centering\includegraphics[width=.85\textwidth]{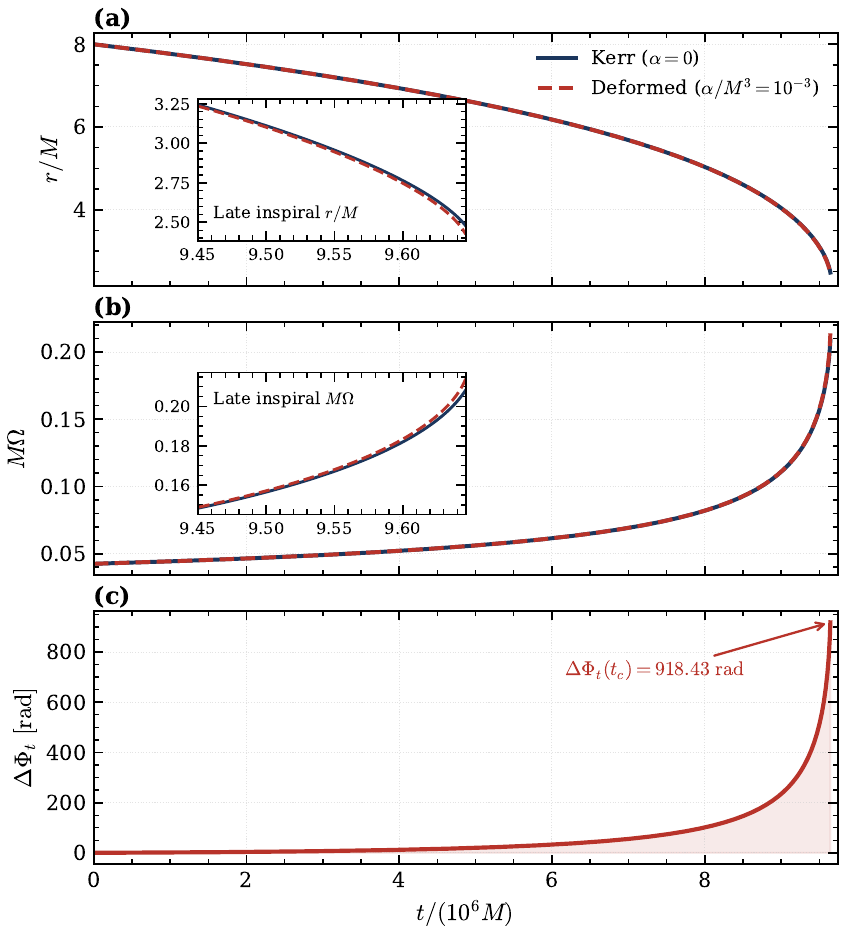}
\caption{Circular trajectories, frequencies, and shared-time phase difference
under the quadrupole closure, ending at the outer cutoff
Eq.~\eqref{eq:stop}. This illustrative closure is distinct from the
fixed-frequency Kerr-flux diagnostic.}\label{fig:trajectory}
\end{figure*}

\begin{figure*}[t]
\centering\includegraphics[width=.95\textwidth]{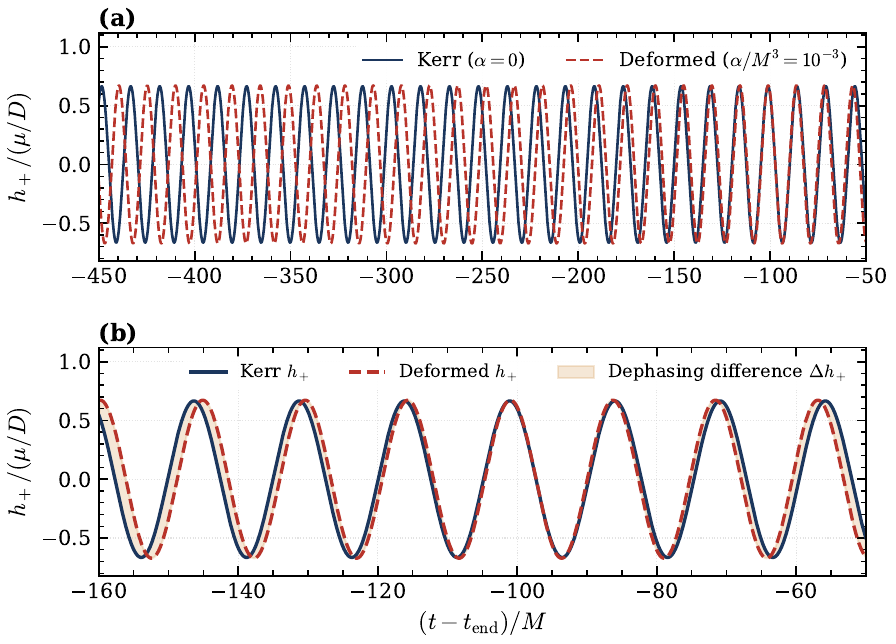}
\caption{Normalized quadrupole waveforms over a late portion of the shared
time interval, with \(\iota=\pi/3\) and the outer termination offset
\(0.1M\). The phase histories use the quadrupole-closure trajectories.
No detector response or parameter fitting is included.}\label{fig:waveform}
\end{figure*}

We use an outer termination for Figs.~\ref{fig:trajectory} and
\ref{fig:waveform},
\begin{equation}
r_{\rm stop}=r_{\rm ISCO}(\alpha)+0.1\quad(M=1).
\label{eq:stop}
\end{equation}
At the deformed endpoint the indicators are
\(\mathcal A_r=1.91\times10^{-4}\) and
\(\mathcal A_\kappa=2.41\times10^{-3}\); the Kerr sequence has
similarly small values. This is a deliberately exterior diagnostic cutoff,
not a dynamically calculated transition radius. On the shared time interval
ending at the earlier stop, the phase difference is
\begin{equation}
\Delta\Phi_t=\Phi_\alpha(t_c)-\Phi_K(t_c)=918.43\ {\rm rad}.
\label{eq:common_time_phase}
\end{equation}
The final frequencies differ in this common-time construction. Its
positive sign should not be compared directly with the negative
fixed-frequency phase in Table~\ref{tab:fixed_frequency}: the
matching conditions differ, and the principal numerical-Kerr
comparison also uses a different radiation prescription.

\begin{table}[tb]
\caption{Termination sensitivity in the quadrupole closure. The offset
\(\delta\) is measured from each geometry's own ISCO. Indicators are
evaluated at the deformed endpoint; phases are compared at the earlier
common stopping time. Large indicators invalidate an adiabatic reading even
when the formal integral converges.}
\label{tab:termination}
\begin{ruledtabular}\begin{tabular}{crrr}
\(\delta/M\)&\(\Delta\Phi_t\) [rad]&\(\mathcal A_r\)&\(\mathcal A_\kappa\)\\\hline
0.1 &918.43&\(1.91\times10^{-4}\)&\(2.41\times10^{-3}\)\\
0.03&960.19&\(5.90\times10^{-4}\)&\(4.76\times10^{-2}\)\\
0.01&965.93&\(1.73\times10^{-3}\)&\(7.34\times10^{-1}\)\\
0.001&966.76&\(1.71\times10^{-2}\)&\(2.31\times10^2\)\\
0.0001&966.77&\(1.71\times10^{-1}\)&\(7.30\times10^4\)\\
\end{tabular}\end{ruledtabular}
\end{table}

Table~\ref{tab:termination} recovers the formal approximately \(967\) rad
value at \(\delta=10^{-4}M\), but explicitly excludes its interpretation
as a controlled near-ISCO waveform. Its formal termination-time difference
is \(-4.51696\times10^3M\); it is not a physical plunge-time prediction.
Convergence of the last digits as \(\delta\) decreases is a numerical fact,
not convergence of the adiabatic approximation. A transition model and a
radiation completion would be required to use this regime physically.

\begin{figure*}[t]
\centering\includegraphics[width=.95\textwidth]{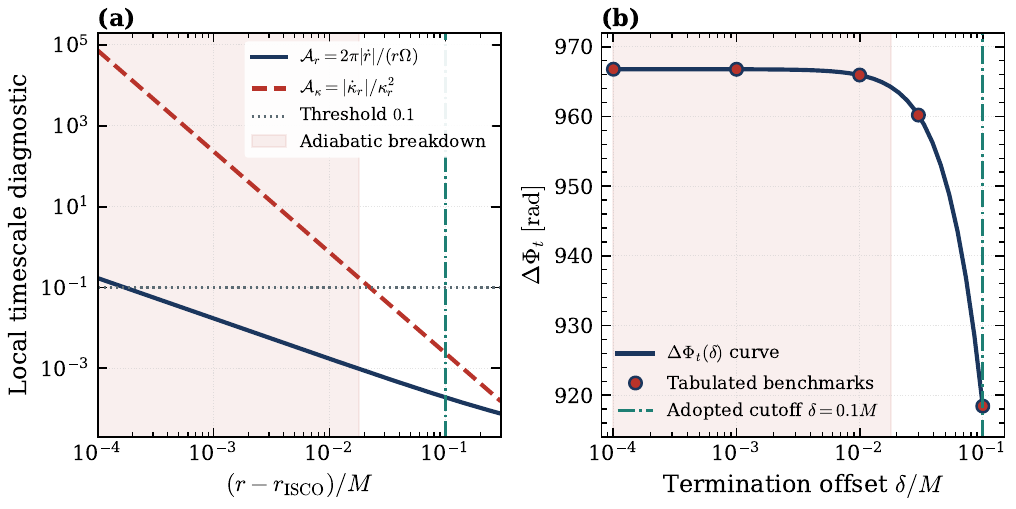}
\caption{Left: local timescale diagnostics at successive circular radii for the
deformed benchmark under the quadrupole closure. The dotted level 0.1 is
a visual guide, not a calibrated error threshold. The radial-epicycle
hierarchy fails well before the formal integral reaches the ISCO. Right:
shared-time phases at the termination offsets listed in
Table~\ref{tab:termination}; apparent saturation of the formal phase does
not establish validity of the approximation.}
\label{fig:adiabaticity}
\end{figure*}

\subsection{Fixed-frequency comparison and flux sensitivity}
The principal conservative comparison uses shared frequency endpoints
\begin{align}
M\Omega_i&=\Omega_K(8)=0.04250360335,\nonumber\\
M\Omega_f&=\Omega_K(3)=0.1640378932.
\label{eq:frequency_endpoints}
\end{align}
In the dimensionless notation, the factors \(M\) simply remind the reader
how frequency is restored. Both branches remain outside the marginal orbits.
The deformed endpoint radii are solved at these same frequencies. For
\(M=10^6M_\odot\), the dominant harmonic covers approximately
\(2.75\)--\(10.60\) mHz.

\begin{table}[tb]
\caption{Changes relative to Kerr at the same frequency endpoints. The flux
is shared at fixed frequency except for the explicit \(b\) templates. No
row supplies a physical non-Kerr radiation calculation.}
\label{tab:fixed_frequency}
\begin{ruledtabular}\begin{tabular}{lrr}
Baseline and template&\(\Delta T/M\)&\(\Delta\Phi_\Omega\) [rad]\\\hline
Quadrupole, \(b=0\)&-2560.37&-460.75\\
Quadrupole, \(b=+1\)&-4165.00&-726.01\\
Quadrupole, \(b=-1\)&-954.80&-195.29\\
Numerical Kerr, \(b=0\)&-2493.03&-435.69\\
Numerical Kerr, \(b=+1\)&-4078.58&-691.13\\
Numerical Kerr, \(b=-1\)&-906.60&-180.07\\
\end{tabular}\end{ruledtabular}
\end{table}

For the numerical Kerr baseline,
\(T_K=9.773619933\times10^6M\) and
\(\Phi_K=1.219126999\times10^6\) rad. The total Kerr-to-quadrupole
flux ratio is approximately 0.9430 at \(r_K=8M\), 0.9749 at \(6M\),
1.0737 at \(4M\), and 1.2118 at \(3M\).
Replacing the baseline changes the frozen-flux conservative phase by
25.06 rad, or 5.44\% relative to the quadrupole value. The sign and
hundreds-of-radians scale persist in this test, while the detailed number is
baseline dependent. This robustness statement is restricted to these two
specified baselines and this interval.

\begin{figure*}[t]
\centering\includegraphics[width=.95\textwidth]{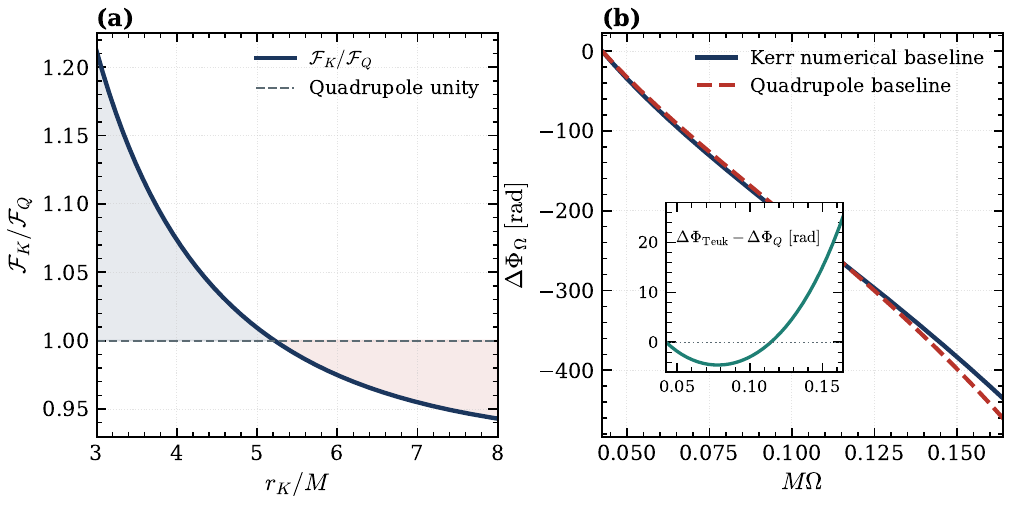}
\caption{Left: tabulated total Kerr flux divided by the quadrupole flux at
the corresponding Kerr radius. Right: fixed-frequency conservative phase
changes for the same metric with each flux frozen between the two geometries.
The difference measures baseline sensitivity, not an environmental flux.}
\label{fig:flux_baseline}
\end{figure*}

The templates in Table~\ref{tab:fixed_frequency} change the numerical-Kerr
result to \(-691.13\) and \(-180.07\) rad. They illustrate that a
fractional flux change no larger than \(10^{-3}\) can contribute hundreds
of radians in these scenarios. They do not bracket a known physical
correction, and cannot support a detector error budget.

\begin{figure*}[t]
\centering\includegraphics[width=.95\textwidth]{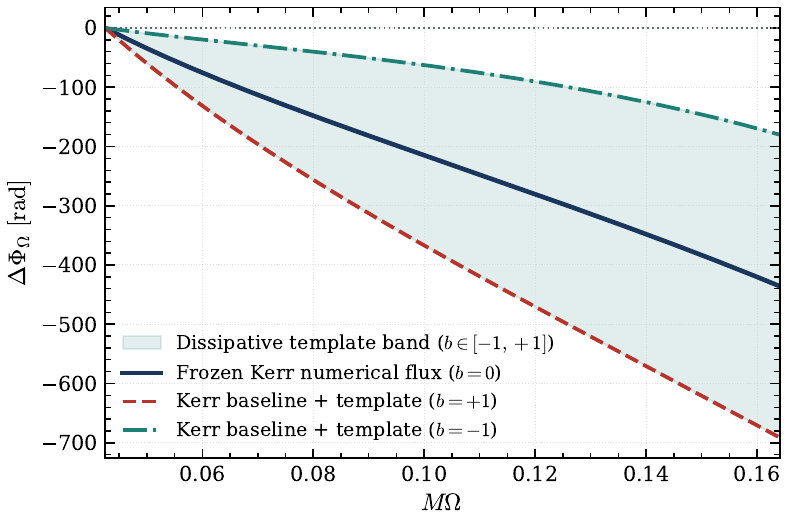}
\caption{Accumulated phase changes with the numerical Kerr baseline and
explicit flux templates. All models have the same energy sequence and shared
frequency endpoints. The templates are scenarios, not confidence bounds.}
\label{fig:frequency_phase}
\end{figure*}

\subsection{Dependence on the metric prescription}
\begin{figure*}[t]
\centering\includegraphics[width=.95\textwidth]{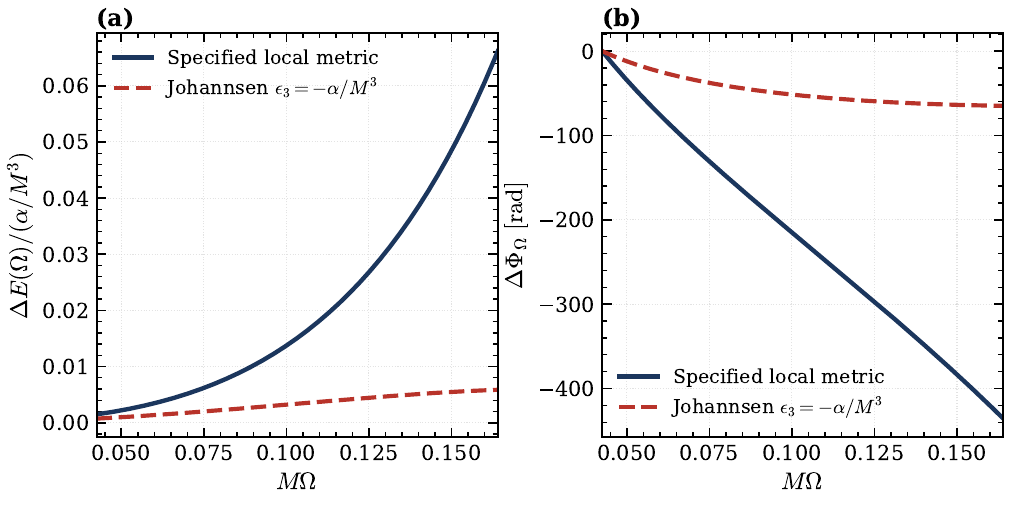}
\caption{Metric-prescription comparison with matched leading asymptotic
\(tt\) coefficients. Both use the same mass, spin, frequency endpoints,
and frozen numerical Kerr flux. The energy-frequency relation and phase
remain sensitive to the other metric components.}\label{fig:metric_comparison}
\end{figure*}

Using Eq.~\eqref{eq:johannsen_comparator} with
\(\epsilon_3=-10^{-3}\) and the same numerical Kerr flux gives
\begin{align}
r^J_{\rm ISCO}/M&=2.320841666,\nonumber\\
M\Omega^J_{\rm ISCO}&=0.225451931,\nonumber\\
\Delta\Phi_\Omega^J&=-64.87\ {\rm rad}.
\label{eq:johannsen_result}
\end{align}
These differ substantially from the local metric despite the matched leading
\(tt\) tail. Figure~\ref{fig:metric_comparison} compares both the
energy-frequency changes and the resulting frozen-flux phases. It shows why
an \(r^{-3}\) correction to one component does not define a universal
EMRI signature. The result is a prescription-dependent benchmark rather than
a claim that this simple family improves on general parameterized metrics.

\subsection{Analytic-kludge waveform illustration}
For the waveform illustration we integrate
\begin{equation}
\phi(t)=\int_0^t\Omega(t')dt',\qquad \Phi_{\rm GW}=2\phi.
\label{eq:wave_phase}
\end{equation}
For a circular trajectory embedded in a flat-space quadrupole prescription,
let \(A_Q=2\mu r^2\Omega_{\rm phys}^2/D\), with physical radius and frequency.
The polarizations can be written as
\begin{align}
h_+&=-A_Q(1+\cos^2\iota)\cos\Phi_{\rm GW},\nonumber\\
h_\times&=-2A_Q\cos\iota\sin\Phi_{\rm GW}.
\label{eq:waveform}
\end{align}
In dimensionless \(M=1\) notation, \(A_Q/(\mu/D)=2r^2\Omega^2\).
These equations state the normalization and polarization convention explicitly;
the orientation sign can be absorbed into a polarization choice.
We set \(\iota=\pi/3\); distance rescales the amplitude only.
The trajectory-to-flat-space identification is part of the kludge, and does
not calculate wave propagation in the deformed geometry.

The kludge includes neither eccentricity, inclination evolution, higher
harmonics, self-force corrections, nor a detector response
\cite{BarackCutler2004,BarackPound2019,PoundWardell2021}. Its amplitude
and phase cannot support precision likelihoods. A visible drift illustrates
the conditional orbital calculation rather than establish detectability.

\section{Interpretation and Requirements for Parameter Inference}
\label{sec:observations}
\subsection{What a phase diagnostic can establish}
The effective coefficient in Eq.~\eqref{eq:full_metric} describes this
particular metric family. The accumulated phase measures a conditional response, not a measurement
forecast. The numerical Kerr flux removes the use of a quadrupole-only vacuum
baseline in the rotating conservative comparison, while the explicit templates
show how sensitive the rotating benchmark is to dissipative corrections
\cite{Cardoso2022,Speeney2024,DattaMaselli2026,RahmanTakahashi2026}. In the
spherical finite-cloud EC calculation of Appendix~\ref{app:finite_cloud}, the
mixed metric--matter perturbations $(\Psi_{\ell m},\Phi_{{\rm f},\ell m})$ and
their radiative and acoustic energy fluxes are computed directly.

\subsection{Detector inner products and parameter degeneracies}
For a detector with one-sided
power spectral density \(S_n(f)\),
\begin{equation}
(h_1|h_2)=4\operatorname{Re}\int_0^\infty
\frac{\tilde h_1(f)\tilde h_2^*(f)}{S_n(f)}df,\qquad
\rho=\sqrt{(h|h)}.
\label{eq:inner_product}
\end{equation}
The inner products of independent detectors add. A reference LISA noise model
is described in Ref.~\cite{RobsonCornishLiu2019}. A mismatch would be
\begin{equation}
\mathcal M=1-\max_{\Delta t,\Delta\phi}
\frac{(h_1|h_2)}{\sqrt{(h_1|h_1)(h_2|h_2)}},
\label{eq:mismatch}
\end{equation}
with an explicit alignment convention. Parameter inference would also vary
mass, central spin, initial orbital parameters, sky position, orientation,
distance, and environmental degrees of freedom. A Fisher matrix,
\begin{equation}
\Gamma_{ij}=(\partial_i h|\partial_j h),
\label{eq:fisher}
\end{equation}
only approximates local inference when its waveform, parameter space, and
high-signal assumptions are appropriate \cite{CutlerFlanagan1994}.

Numerical SNRs, mismatches, Fisher errors, and confidence contours require
validated waveforms and inference settings that are not supplied by the
present closure. In particular, a contour
in an \((\alpha,\rho_0)\) plane assumes a microphysical map that has not
been derived. No quantitative detector or parameter-estimation bounds are reported.

A limited shape test makes the parameter issue concrete. At 201 uniformly
spaced frequencies in Eq.~\eqref{eq:frequency_endpoints}, project
\(\Delta E(\Omega)\) onto the Kerr directions
\(E_{,a_*}\) and \(\Omega E_{,\Omega}\), where the second corresponds
to a small mass change at fixed physical frequency and fixed dimensionless
spin. The least-squares coefficients are
\(\delta a_*=-4.34\times10^{-4}\) and
\(\delta\ln M=8.68\times10^{-5}\). The remaining relative Euclidean
norm of the energy change is 2.69\%. Thus much of this particular energy
shape can be mimicked by small Kerr-parameter changes. This unweighted
energy-space calculation includes neither flux changes nor detector noise,
and is not a mismatch, likelihood, or observational constraint.

\subsection{Microphysical interpretation and scale separation}
The orbital calculation does not determine a dark-matter density,
polarization, or fundamental torsion coupling. Such an interpretation
requires a microscopic matter action and a background solution that
reproduces all relevant metric components with matched boundary and
normalization conditions. The averaged-fluid coefficient \(C\) and its
dimensionless counterpart \(\zeta\) in the appendices are not identified
with \(\alpha\). In particular, a single spin-density estimate cannot
supply that map or a quantitative suppression of this metric's coefficient.

The calculated local sensitivity with the numerical Kerr flux frozen is
\begin{equation}
\left.\frac{\partial\Delta\Phi_\Omega}{\partial(\alpha/M^3)}\right|_0
\simeq-4.3573\times10^5\ {\rm rad}.
\label{eq:linear_coefficient}
\end{equation}
This derivative quantifies the response of the prescribed rotating geometry
under the stated endpoint and flux convention. In the spherical EC sector of
Appendix~\ref{app:finite_cloud}, where the simultaneous coupled orbit-phase
system $(r_{\rm orb}(v),\Phi_{\rm GW}(v))$ is integrated over the finite-cloud
evolution window $v\in[0,200M_0]$ using the exact circular-orbit identity
$\partial_r E=\Omega\partial_r L$, the cloud-embedded orbit contracts by
$\Delta r_{\rm orb}(200M_0)=-8.233\times10^{-4}M_0$ (versus
$\Delta r_{\rm vac}(200M_0)=-9.865\times10^{-5}M_0$ in vacuum) and accumulates
a genuine time-dependent phase shift
$\Delta\Phi_{\rm GW}^{\rm TD}(200M_0)=+1.367\times10^{-3}\text{ rad}$ at
$\eta=10^{-5}$. (A hypothetical static-slice integral across
$r_{\rm vac}\in[6.1M_0,8.0M_0]$ on the frozen $v=0$ profile would give
$-0.53222\text{ rad}$ conservative and $+0.46609\text{ rad}$ dissipative, or
$-0.06614\text{ rad}$ net, but because the finite cloud accretes on
$v\sim200M_0\ll T_{\rm rad}$, only the coupled $v\in[0,200M_0]$ integration
represents a physical finite-cloud phase prediction.) No detector constraint on
fundamental minimal torsion is claimed.

\subsection{Environmental and modeling systematics}
Ordinary matter gravity changes the energy-frequency relation, whereas drag,
accretion, and matter excitations can change the balance law. In
collisionless-spike models, the secondary can also redistribute the dark
matter through dynamical-friction feedback \cite{Kavanagh2020}; that
mechanism differs from the perfect-fluid inflow of
Appendix~\ref{app:ec_accretion}. Central spin
and the effective deformation can both influence the late circular sequence.
Self-force terms and detector modulation add further correlated waveform
structure. These effects motivate a joint source, perturbation, and inference
model; their relative importance cannot be ranked quantitatively by the
present metric alone. This restriction preserves the useful phenomenological
calculation while identifying the additional work needed to connect it to
observations.

\section{Conclusions and Discussion}
\label{sec:conclusions}
We have studied the circular equatorial response of a specified near-zone
Kerr deformation, retaining spin matter as motivation while making no claim
of an Einstein--Cartan source solution. The four-dimensional definition is
regular on the demonstrated exterior domain. It is neither a unique angular
completion nor a validated black-hole interior or horizon extension.

The benchmark ISCO displacement is \(3.6538\times10^{-3}M\), accompanied
by \(M\Delta\Omega_{\rm ISCO}=-3.9168\times10^{-4}\).
On a shared frequency interval, using numerical Kerr perturbation-theory
fluxes as a frozen baseline gives \(-435.69\) rad of conservative
dephasing. A quadrupole baseline gives \(-460.75\) rad, so the diagnostic
changes by 5.44\% under this replacement. A Johannsen comparison matched
in its leading \(tt\) tail gives \(-64.87\) rad under the same flux,
showing that the other metric components materially affect the response.

Local timescale checks separate an exterior adiabatic illustration from an
uncontrolled near-ISCO extrapolation. With the quadrupole closure, a cutoff
\(0.1M\) outside the marginal orbit gives 918.43 rad at shared time.
The formal approximately 967 rad value at \(10^{-4}M\) is numerically
reproducible but lies in a regime with a failed radial-epicycle hierarchy;
it is not a controlled waveform or plunge prediction.

Unknown leading flux changes can contribute at the same
\(O(\alpha/\eta)\) order as conservative changes. Explicit flux scenarios
and the Kerr-parameter energy-shape projection show that radiation and
parameter degeneracies remain. The calculation therefore supports neither
detectability nor dark-matter or fundamental-coupling bounds. A physical
extension requires a specified source and perturbation theory; observational
inference additionally requires a validated waveform and detector model.

The independent spherical statistical-spin test in
Appendix~\ref{app:ec_source} makes part of that source problem explicit.
Its effective fluid is variationally and thermodynamically closed, and
local equilibrium solutions satisfy the Einstein and conservation equations.
However, the cold decreasing spike fails hydrostatic balance on the stated
dilute branch, while the pressure-supported diagnostic solutions have no
regular inner completion. This is a restriction on the chosen static
isotropic closure, not a general exclusion of EC matter around black holes.
It also prevents these local solutions from supplying the inner boundary
condition for a physical EMRI radiation calculation.

Allowing isotropic radial accretion avoids that static obstruction.
Appendix~\ref{app:ec_accretion} supplies a transonic statistical-spin flow,
a positive matter energy inflow, and a first metric correction including
black-hole mass growth, with the expected quadratic Einstein remainder. This
weak-backreaction inflow solution provides the transonic seed profile that is
completed into a globally regular, asymptotically flat finite cloud in
Appendix~\ref{app:finite_cloud}.

Appendix~\ref{app:finite_cloud} completes the six-state coarse-grained
Einstein--Cartan spin-bundle analysis and the coupled time-dependent EMRI
radiation calculation on the nonlinear finite cloud with a vacuum exterior.
Using an orthonormal comoving tetrad whose spatial triad is Fermi--Walker
transported along the accelerating radial flow, direct four-dimensional
coordinate contraction confirms $\tfrac12 s_{\mu\nu}s^{\mu\nu}=\chi_s n^2$ and
exact 4D MPD transport for each branch, while proving that spherical isotropy
requires the local six-state coarse-grained ensemble ($\langle s_{\mu\nu}\rangle=0$,
$\langle B_{\mu\nu}\rangle=0$, $\langle Q_{\mu\nu}\rangle=\tfrac23\chi_s n^2 h_{\mu\nu}$)
because an individual polarized branch produces anisotropic quadratic spin
stress and off-diagonal Belinfante divergences. Solving the simultaneous
4-component BVP for even-parity Zerilli--Moncrief and relativistic acoustic
enthalpy perturbations across all polar modes and summing all 9 multipole
pairs for $2\le\ell\le4$
\cite{Zerilli1970,Chandrasekhar1983,LindblomDetweiler1983,HopperEvans2010,Blanchet2014}
yields $\mathcal F_E^{\rm tot}/(\mu/M_0)^2=1.94959787\times10^{-4}$ at
$M_0\Omega=8^{-3/2}$ ($84.54\%$ from $(2,\pm2)$ and $2.34\%$ from $\ell=4$),
with a paired $(2,2)$ EC-minus-zero-spin fractional shift of
$-1.4945\times10^{-8}$ (boundary drift $9.05\times10^{-15}$) and a $+2.53\%$
EC enhancement of the total polar acoustic matter flux. Over $v\in[0,200M_0]$,
the quasi-stationary source flux shift grows from $-1.879\times10^{-6}$ to
$+3.013\times10^{-5}$; evolving the $(1+1)$D Eddington--Finkelstein wave PDE
accounts for the causal null-cone delays $\Delta v_{\rm prop}=92.32M_0,154.26M_0,235.92M_0$
to $r_{\rm ext}=50M_0,80M_0,120M_0$, giving extracted flux shifts of
$+1.557\times10^{-5}$ at $r_{\rm ext}=50M_0$ ($v_{\rm obs}=200M_0$) and
$+2.956\times10^{-5}$ at $r_{\rm ext}=120M_0$ at retarded arrival
$v_{\rm obs}=435.92M_0$, while simultaneous integration of
$(r_{\rm orb}(v),\Phi_{\rm GW}(v))$ yields a coupled orbital contraction
$\Delta r_{\rm orb}(200M_0)=-8.233\times10^{-4}M_0$ and time-dependent phase
shift $\Delta\Phi_{\rm GW}^{\rm TD}(200M_0)=+1.367\times10^{-3}\text{ rad}$
for $\eta=10^{-5}$.

\begin{acknowledgments}
This work made use of numerical Kerr flux data provided by S.~A.~Hughes
for Ref.~\cite{Taracchini2014}, hosted as part of the Black Hole
Perturbation Toolkit at \url{https://bhptoolkit.org}.
\end{acknowledgments}

\appendix
\section{Numerical verification and reproducibility}
\label{app:verification}
The companion source archive includes the calculation, figures, frequency-phase
table, and machine-readable numerical results. No original numerical
program or trajectory data were present in the supplied source archive;
the diagnostics were rebuilt from its stated metric and flux equations.

At \(\alpha=0\), a 201-point sample of \(2.4\leq r\leq8\) reproduces
the analytic Kerr \(\Omega,E,L_z\) to maximum absolute errors
\(5.6\times10^{-17},5.6\times10^{-16},2.7\times10^{-15}\), respectively.
The radial potential gives the independent marginal-stability residuals
reported with Eq.~\eqref{eq:isco_numeric}. The symbolic substitution
residual for the conditional static integral vanishes identically.

For completeness, vertical perturbations of the equatorial circular orbit have
coordinate-time frequency
\begin{equation}
\Omega_\theta^2=-\frac{1}{2g_{\theta\theta}}
\left[\partial_\theta^2g_{tt}
+2\Omega\partial_\theta^2g_{t\phi}
+\Omega^2\partial_\theta^2g_{\phi\phi}\right]_{\theta=\pi/2}.
\label{eq:vertical_stability}
\end{equation}
It is positive on a 1001-point sample between each radial marginal orbit
and \(r=8\) for \(\alpha=-10^{-3},0,10^{-3}\); the smallest sampled
value is \(1.5877\times10^{-3}\) in \(M^{-2}\) units. Thus no vertical
instability is found on the sampled branches. This is a numerical branch
check, not a generic-inclination integrability claim.

The numerical Kerr flux file contains 10,000 rows in radius, with separate
infinity and horizon kernels. The companion archive preserves the downloaded
table, source description, retrieval date, and SHA-256 checksum. Sorting in
radius is the only rearrangement; no data point is removed. The weak-field
infinity-flux ratio to \((32/5)r^{-5}\) approaches unity, checking the
mass normalization. The total positive flux is interpolated cubically in
\((\ln r,\ln f)\), without extrapolation outside the tabulated domain.

Every source-table interval within the frequency range is included in a
composite Gauss--Legendre integral. This prevents narrow features in the
input from being missed by a global adaptive quadrature. Four- and
eight-point interval rules agree in the conservative phase to below
\(10^{-7}\) rad. A held-out alternate-node reconstruction has a maximum
fractional discrepancy \(1.72\times10^{-5}\) near \(r=3.13818M\),
where the source contains a localized flux feature; its 99th-percentile
discrepancy is below \(5\times10^{-12}\). Keeping all rows versus using
the thinned reconstruction changes the phase difference by approximately
\(3.3\times10^{-6}\) rad. A shape-preserving interpolant gives a
consistent integral. These tests measure interpolation sensitivity, not the
original flux solver's uncertainty or non-Kerr radiation error.

The local linear coefficient with the numerical Kerr baseline is checked
by both signs at \(\alpha=10^{-5}\). Its finite-\(\alpha=10^{-3}\)
remainder in the conservative phase is approximately 0.035 rad. Independent
radial quadrature and sampled monotone interpolation check the common-time
trajectory illustration. Table~\ref{tab:termination} separately records
termination dependence and local timescale failure; neither is replaced by
an integration-tolerance test.

The vertical frequency in Eq.~\eqref{eq:vertical_stability} applies to
the chosen angular completion. A smooth addition
Eq.~\eqref{eq:angular_freedom} changes it while preserving the equatorial
orbit sequence. At the equator, the local spin two-form
\(s_{\hat r\hat\phi}=\sigma\) has spatial dual
\(S^{\hat\theta}=-\sigma\); since \(\hat\theta\) points along
\(-z\) there, the illustrative spin arrows in Fig.~\ref{fig:scope}
are drawn along \(+z\). This orientation check is not a solved global
polarized-fluid configuration.

\section{A separate spherical, statistically unpolarized EC source test}
\label{app:ec_source}

This appendix tests a different and more restrictive matter model than the
polarized motivation in Fig.~\ref{fig:scope}. It does not identify the
prescribed rotating metric with an EC solution. We first close an averaged
spin-fluid model, then test its equilibrium and boundary conditions. The
result is a set of locally self-consistent static equilibria and conditional
obstructions to completing them as regular static black-hole atmospheres.
Those static obstructions motivate the dynamical inflow of
Appendix~\ref{app:ec_accretion} and the completed nonlinear finite-cloud
evolution, coarse-grained six-state Kramers-bundle spin-transport closure, and coupled
$(\ell,m)=(2,2)$+matter EMRI radiation calculation in
Appendix~\ref{app:finite_cloud}.

\subsection{Statistical closure and bulk equations}

We use signature $(-+++)$, $G=c=1$, $\kappa=8\pi$, and
$Q^\lambda{}_{\mu\nu}=\Gamma^\lambda{}_{[\mu\nu]}$. The microscopic
Weyssenhoff spin tensor satisfies $s_{\mu\nu}u^\nu=0$. Its first moment vanishes,
but its invariant variance is prescribed to be
\begin{align}
 \langle s_{\mu\nu}\rangle&=0,\qquad C\equiv\kappa\chi_s>0,\nonumber\\
 {\cal S}^2=\tfrac12\langle s_{\mu\nu}s^{\mu\nu}\rangle
 &=\chi_s n^2.
 \label{eq:ec_statistics}
\end{align}
Here $n$ is proper particle number density and $\chi_s$ is a fixed
spin-statistics parameter, not a net polarization. We adopt locally
isotropic spin statistics in the rest frame, with
$\langle s_{ij}s_{kl}\rangle=({\cal S}^2/3)
(\delta_{ik}\delta_{jl}-\delta_{il}\delta_{jk})$.
The torsion is eliminated before averaging its quadratic contribution.
Replacing the spin tensor by its zero mean before this elimination would
incorrectly remove that contribution.

With the spin normalization of Ref.~\cite{Brechet2007}, the effective
energy density and isotropic pressure in this closure are
\begin{align}
 \rho_e(n)&=\rho(n)-Cn^2, &p_e(n)&=p(n)-Cn^2,\nonumber\\
 p(n)&=n\rho_{,n}-\rho, &
 T^e_{\mu\nu}&=(\rho_e+p_e)u_\mu u_\nu+p_e g_{\mu\nu}.
 \label{eq:ec_effective_eos}
\end{align}
Linear spin-divergence terms in the unaveraged effective tensor vanish
under the assumed local average. Neglect of connected metric--spin
correlations and an averaging scale smaller than the background variation
scale are additional closure assumptions. In the static setting alone, Eq.~\eqref{eq:ec_statistics} enters as a
statistical closure assumption; in Appendix~\ref{app:finite_cloud}
[Eqs.~\eqref{eq:microscopic_tetrad}--\eqref{eq:exact_mpd_transport}], we
construct the explicit six-state Kramers-bundle spin tensor on a
Fermi--Walker-transported comoving tetrad that satisfies four-dimensional MPD
spin transport identically and proves that while a single polarized branch
breaks spherical isotropy via off-diagonal Belinfante divergences and
anisotropic quadratic spin stress, the local six-state coarse-grained
ensemble yields both Eq.~\eqref{eq:ec_statistics} and its linear perturbation.

The reduced bulk model has the variational perfect-fluid description
$S_e=-\int\sqrt{-g}\,\rho_e(n)\,d^4x$, with conserved particle current
and isentropic fluid variations \cite{Brown1993}. It is thermodynamically consistent:
$p_e=n\rho_{e,n}-\rho_e$. Particle conservation gives
$\dot n=-n\Theta$; the Weyssenhoff spin propagation law gives
$\dot{\cal S}=-{\cal S}\Theta$ in this closure
\cite{Brechet2007}. Thus a constant spin variance per squared particle
density is consistent with transport. This specifies a reduced matter
model, not a unique prediction of a microscopic dark sector. Its coefficient
$C$ has no established relation to the metric parameter $\alpha$.

For
\begin{align}
 ds^2&=-N^2dt^2+B^{-1}dr^2+r^2d\Omega_2^2,\nonumber\\
 N&=e^\nu,\qquad B=1-2m/r,
\end{align}
the effective perfect-fluid equations have the usual TOV structure
\cite{Tolman1939,OppenheimerVolkoff1939}, with the spin-corrected
density and pressure as sources:
\begin{align}
 m'&=4\pi r^2\rho_e,\nonumber\\
 \nu'&=\frac{m+4\pi r^3p_e}{r(r-2m)},\nonumber\\
 p_e'&=-(\rho_e+p_e)\nu'.
 \label{eq:ec_tov}
\end{align}
The angular Einstein equation then follows identically. A static fluid
has $u^\mu=N^{-1}(1,0,0,0)$ and $\Theta=0$, but its acceleration is
$a^r=B\nu'$, generally nonzero. Hydrostatic support must not be removed
by equating static flow with geodesic flow.

Define the effective enthalpy per particle
\begin{equation}
 h_e=\rho_{e,n}=\frac{\rho_e+p_e}{n}.
\end{equation}
Since $dp_e=n\,dh_e$, Eq.~\eqref{eq:ec_tov} has the first integral
\begin{equation}
 h_e N={\cal H},\qquad {\cal H}>0
 \label{eq:ec_first_integral}
\end{equation}
on a nonvacuum positive-enthalpy branch. Equations
\eqref{eq:ec_effective_eos}--\eqref{eq:ec_first_integral}, together
with boundary data and an ordinary equation of state, close the static
bulk problem. They do not permit an independently imposed density profile.

\subsection{Cold spike test and boundary obstructions}

To test the original cold-spike motivation in the new statistical model,
take $\rho=m_\chi n$, $p=0$. The spin-corrected pressure is negative,
$p_e=-Cn^2$. Hydrostatic balance requires, for $n>0$,
\begin{equation}
 n'=\frac{m_\chi-2Cn}{2C}\nu'.
 \label{eq:ec_cold_gradient}
\end{equation}
On the dilute positive-enthalpy branch $2Cn<m_\chi$ and in a region
where $m+4\pi r^3p_e>0$, density must increase outward. It therefore
cannot be the assumed decreasing spike. More explicitly, substitution of
$n=A r^{-3/2}$ into the conservation equation leaves the residual
\begin{equation}
 \mathcal R=p_e'+(\rho_e+p_e)\nu'
 =\frac{3Cn^2}{r}+n(m_\chi-2Cn)\nu'>0.
 \label{eq:ec_cold_residual}
\end{equation}
This statement uses both stated inequalities; it is not an exclusion of
all strong-spin or anisotropic configurations. The local isentropic
compressibility gives the additional diagnostic
\begin{equation}
 c_e^2=\frac{p_{e,n}}{\rho_{e,n}}
 =-\frac{2Cn}{m_\chi-2Cn}<0.
 \label{eq:ec_cold_sound}
\end{equation}
Within the assumed fluid regime, this is a local gradient instability,
not a calculated global mode spectrum. Neither the single-component spin
ansatz nor its invariant normalization can replace these conservation and
stability tests. Related, but distinct, restrictions on pressure-free
spin-supported stars were proved in Ref.~\cite{LuzCarloni2019}; that
stellar proposition is not used as a theorem about black-hole halos here.

There is also an equation-of-state-independent obstruction for ordinary
effective stresses. Suppose a connected static fluid occupies
$r_{\rm in}<r<r_{\rm out}$, with $m>0$, $r>2m$, $\rho_e\geq0$,
$p_e\geq0$, and $\rho_e+p_e>0$ wherever matter is present. Assume
both boundaries join vacuum without a surface stress layer. At the level
of the averaged Einstein equations, continuity of the induced metric and
extrinsic curvature \cite{Israel1966} requires $p_e(r_{\rm in})=0$.
Microscopic EC junction conditions are not established by this effective
matching argument. However,
Eq.~\eqref{eq:ec_tov} gives $p_e'<0$ throughout the matter region.
Integration outward from the inner boundary would give $p_e<0$,
contradicting the assumptions. Thus an unsupported finite, hollow,
isotropic atmosphere with these properties is impossible. An arbitrary
inner cutoff does not solve this junction problem. Negative effective
pressure, anisotropy, material surface layers, or time dependence lie
outside this argument and require their own models.

Letting the static fluid extend to a horizon is a separate issue. For a
regular horizon $N\to0$, Eq.~\eqref{eq:ec_first_integral} requires
$h_e\to\infty$ if ${\cal H}>0$. A branch with finite maximum enthalpy
cannot reach such a horizon. For an ordinary zero-spin polytrope,
$h_e\to\infty$ instead entails divergent density and pressure; that
does not provide a finite-stress horizon. These are statements about the
chosen static positive-enthalpy branch, not about dynamical EC collapse.

\subsection{Local solutions with an explicit pressure law}

To test whether ordinary pressure repairs the cold calculation, choose
the simple isentropic polytrope
\begin{equation}
 \rho=m_\chi n+\frac{K n^\Gamma}{\Gamma-1},\qquad
 p=K n^\Gamma,\qquad \Gamma=5/3.
 \label{eq:ec_polytrope}
\end{equation}
This is an illustrative fluid equation of state. Its high-density
continuation is not identified with a nonrelativistic microscopic Fermi
gas. The exponent retains a useful connection with the original profile:
in a central-mass-dominated Newtonian limit, neglecting spin and setting
the enthalpy integration constant to zero gives
$n\propto r^{-1/(\Gamma-1)}=r^{-3/2}$. Ordinary pressure, rather than
the spin term alone, supplies that limiting hydrostatic support. The
profile need not survive in the relativistic inner region or near a free
outer surface.

Set $x=n/n_*$, $\varrho_*=m_\chi n_*$,
$q=K n_*^\Gamma/\varrho_*$, and
$\zeta=C n_*^2/\varrho_*$. Then
\begin{align}
 \frac{\rho_e}{\varrho_*}
 &=x+\frac{q x^\Gamma}{\Gamma-1}-\zeta x^2,\nonumber\\
 \frac{p_e}{\varrho_*}&=q x^\Gamma-\zeta x^2,\nonumber\\
 \frac{h_e}{m_\chi}
 &=1+\frac{\Gamma q}{\Gamma-1}x^{\Gamma-1}-2\zeta x.
 \label{eq:ec_dimensionless_eos}
\end{align}
Our diagnostic choices are $\varrho_*M_0^2=10^{-8}$, $q=0.02$,
$r_*=8M_0$, $m(r_*)=M_0$, $x(r_*)=1$, and
$\zeta=0,10^{-3},10^{-2}$. Here $M_0$ is the enclosed mass at the
reference radius. It has not been established as a central black-hole
mass. The values of $\zeta$ are mathematical test parameters; no particle
mass, abundance, or microscopic EC coupling calibration is asserted.

For $\zeta>0$, the first positive-density zero of $p_{e,x}$ is
\begin{equation}
 x_c=\left(\frac{\Gamma q}{2\zeta}\right)^{1/(2-\Gamma)}.
 \label{eq:ec_critical_density}
\end{equation}
Here $h_e$ attains its finite maximum on the positive-compressibility
branch. If $\nu'>0$ and $\rho_e+p_e>0$, the TOV pressure gradient is
nonzero at $x_c$ although $p_{e,x}=0$. Consequently $dx/dr$ diverges.
Changing the integration variable to $\ln x$ allows this endpoint to be
located without mistaking numerical stiffness for a horizon. Continuing
the inverse solution beyond it would turn $r(x)$ and enter a branch with
negative $c_e^2$; it would not extend the same stable atmosphere inward.

\begin{table}[t]
\caption{Endpoints of the local spin-fluid branches. Both have finite
positive effective density and pressure. These are finite-density
compressibility endpoints, not horizons, ISCOs, or measured critical
dark-matter densities.}
\label{tab:ec_endpoints}
\begin{ruledtabular}
\begin{tabular}{ccccc}
$\zeta$ & $x_c$ & $r_c/M_0$ & $m_c/M_0$ & $B_c$\\
\hline
$10^{-3}$ & $4629.630$ & $2.052766$ & $0.999435$ & $0.026255$\\
$10^{-2}$ & $4.629630$ & $7.321154$ & $0.999990$ & $0.726822$
\end{tabular}
\end{ruledtabular}
\end{table}

\begin{figure*}[t]
\centering
\includegraphics[width=\textwidth]{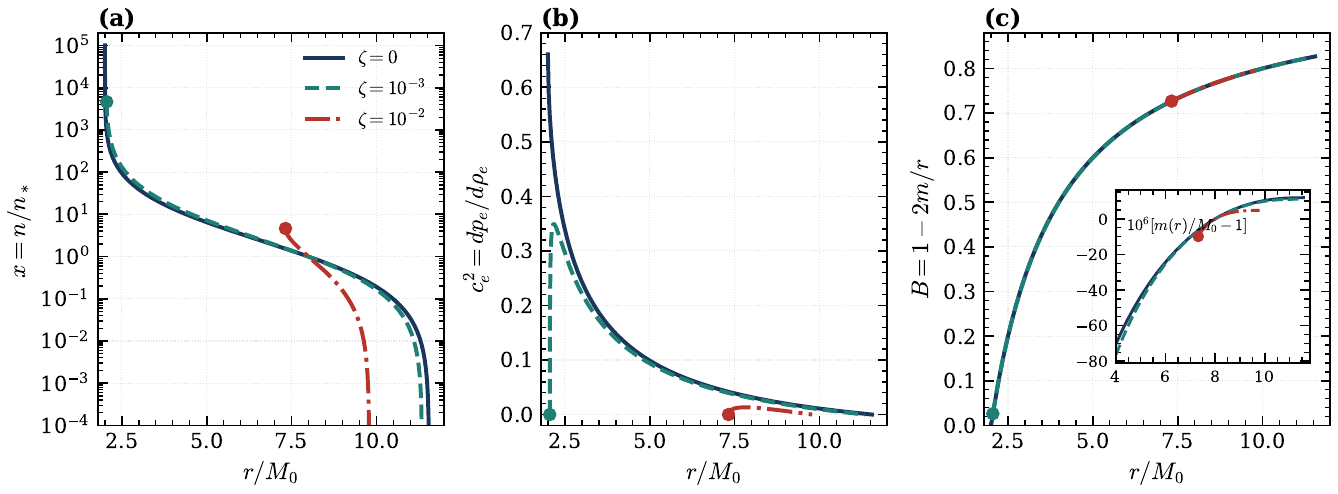}
\caption{Locally self-consistent solutions of the averaged EC TOV system
for the illustrative polytropic closure. Left: particle density. Middle:
effective sound speed squared. Right: $B=1-2m/r$; the nearly overlapping
curves do not imply identical source models. Colored dots mark the
finite-density endpoints for nonzero spin coupling. The zero-spin curve
is stopped at $x=10^5$ only as a diagnostic; it is not a demonstrated
regular horizon solution. Every curve has the same reference data at
$r=8M_0$. None is a completed black-hole atmosphere or a source for the
rotating metric in the main text.}
\label{fig:ec_local}
\end{figure*}

The nonzero-spin endpoints are listed in
Table~\ref{tab:ec_endpoints}. In particular, $B_c>0$ in both cases.
Finite density and pressure at the endpoint do not establish a curvature
singularity; the failure is that of the selected single-valued hydrostatic
branch and its positive compressibility. Toward larger radii, the density
approaches zero at $r/M_0\simeq11.55229,11.34605,9.78832$ for the three
couplings, respectively. The outer limit permits matching to a
Schwarzschild exterior with mass $m(r_{\rm out})$ and fixes the lapse
normalization. It supplies no inner vacuum junction or regular horizon.

Verification uses two different coordinates for integration, direct
Christoffel/Ricci evaluation of the metric, and the analytic test-fluid
first integral on Schwarzschild. The tensor identities and thermodynamic
identity vanish symbolically. On well-conditioned subintervals, the
maximum finite-difference Einstein residual is below
$2\times10^{-11}M_0^{-2}$ and the conservation residual below
$4\times10^{-20}M_0^{-3}$; finite-difference cancellation limits the
angular check. The two integration coordinates agree within
$2\times10^{-14}$ in the reported dimensionless variables. A simultaneous
tighter tolerance and smaller maximum integration step changes each
listed endpoint radius by less than $10^{-11}M_0$. In the test-fluid
limit $\varrho_*M_0^2=10^{-15}$, the numerical density at $r=4M_0$
differs from the analytic zero-self-gravity value by
$9.5\times10^{-12}$ fractionally, including finite self-gravity.
These checks establish numerical consistency of the local reduced
equations, not physical completeness of the atmosphere.

\subsection{Consequences for the radiation calculation}

Even a completed static metric would not alone specify its EMRI energy
flux. If the statistical closure is retained under perturbations, with
no induced polarization or spin-correlation degrees of freedom, the
matter variations include
\begin{align}
 \delta\rho_e&=(\rho_{,n}-2Cn)\delta n,\nonumber\\
 \delta p_e&=(p_{,n}-2Cn)\delta n,\qquad
 c_e^2=\frac{p_{,n}-2Cn}{\rho_{,n}-2Cn}.
 \label{eq:ec_perturbative_eos}
\end{align}
As proved in Appendix~\ref{app:finite_cloud} by linearizing the exact
four-dimensional MPD spin-transport equation~\eqref{eq:exact_mpd_transport}
around the Kramers-doublet background, $\delta(\mathcal S^2/n^2)=0$ holds
along perturbed fluid worldlines in the absence of non-gravitational
polarizing torques, so Eq.~\eqref{eq:ec_perturbative_eos} follows directly from
microscopic spin transport.

With $g_{\mu\nu}\to g_{\mu\nu}+h_{\mu\nu}$ and an unspinning point
secondary, the bulk equations to linear order in its mass are
\begin{align}
 \delta G_{\mu\nu}[h]
 &=8\pi\delta T^e_{\mu\nu}[h,\delta n,\delta u]
   +8\pi T^{\rm pp}_{\mu\nu},\nonumber\\
 \nabla_\mu(\delta n\,u^\mu+n\delta u^\mu)
 &+\tfrac12 n u^\mu\nabla_\mu h=0,
 \label{eq:ec_linear_system}
\end{align}
where $h=g^{\mu\nu}h_{\mu\nu}$, together with the linearized effective
Euler equation and velocity normalization. For covariant velocity
perturbations the stress variation is
\begin{align}
 \delta T^e_{\mu\nu}={}&(\delta\rho_e+\delta p_e)u_\mu u_\nu
 +(\rho_e+p_e)(\delta u_\mu u_\nu+u_\mu\delta u_\nu)
 \nonumber\\
 &+\delta p_e\,g_{\mu\nu}+p_e h_{\mu\nu}.
\end{align}
Thus the explicit spin correction enters both the background and the
matter response. It is not reproduced by inserting a new orbital
frequency into a vacuum Kerr flux. In minimal EC theory the algebraic
torsion does not provide an independent propagating vacuum wave channel;
this does not eliminate the fluid perturbations. Relativistic
environmental radiation calculations likewise require the coupled metric
and matter problem \cite{Cardoso2022,Speeney2024,DattaMaselli2026}.

A completed calculation would impose an ingoing condition at a regular
black-hole horizon, an outgoing condition at infinity, and the metric
junction and Lagrangian pressure condition $\Delta p_e=0$ at a free
material surface. Matter energy transport must also enter the balance
law if such a channel is open. Because the static branches above encounter a finite-density compressibility
endpoint rather than a regular horizon, they cannot supply the inner boundary
condition for the radiation problem. Allowing radial inflow in
Appendices~\ref{app:ec_accretion} and~\ref{app:finite_cloud} removes this
static obstruction, provides a regular horizon and vacuum exterior, and
enables the complete coupled $(\ell,m)=(2,2)$ polar, axial, and matter
perturbation flux calculation presented in Appendix~\ref{app:finite_cloud}.

\section{Transonic inflow and first metric backreaction}
\label{app:ec_accretion}

The failure of the static branches does not exclude radial inflow.
We therefore retain the functional form of the averaged isotropic
spin-fluid closure of
Appendix~\ref{app:ec_source}, but allow $u^r<0$. This changes the
physical model: it is an open accreting system, not a stationary dark-matter
spike. We first solve the transonic test-fluid equations and then compute
their first metric backreaction. This construction solves for the first weak metric backreaction of a
leading conserved source; it is not an exact nonlinear EC solution or
a source for the rotating phenomenological metric.
The hydrodynamic equation of state is also an assumption: the relaxation
scale needed to describe a dark component as this fluid has not been
established for a collisionless dark-matter halo.

\begin{figure*}[t]
\centering
\includegraphics[width=0.88\textwidth]{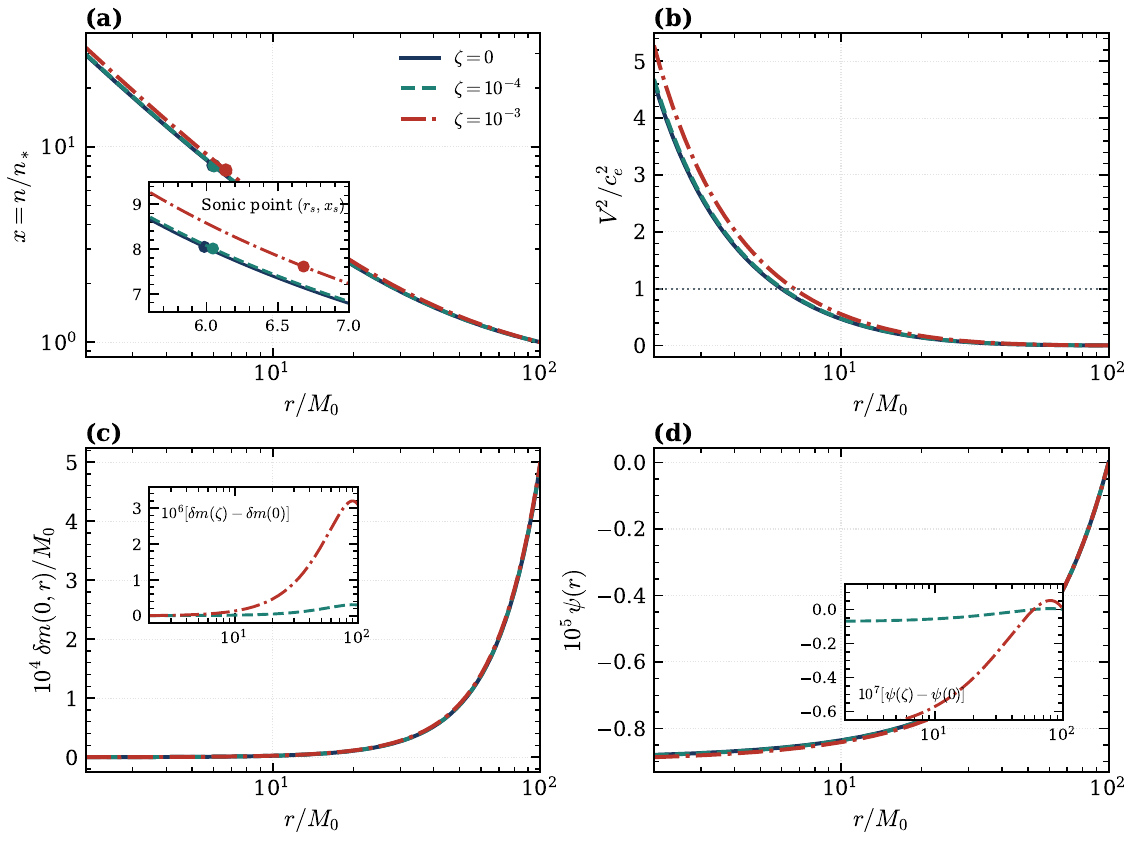}
\caption{Transonic statistical-spin inflow and first metric backreaction.
Top left: density; dots mark sonic points. Top right: squared local radial
speed divided by $c_e^2$, with $V^2=w^2/(f+w^2)$, defined relative to
static Schwarzschild observers outside the horizon and shown with its
horizon limit. Bottom: radial mass correction at $v=0$ and ingoing lapse
exponent. The outer boundary supplies incoming matter. These curves do
not describe the polarized schematic in Fig.~\ref{fig:scope} or the
main-text rotating geometry.}
\label{fig:ec_accretion}
\end{figure*}

\subsection{A regular inflow through the sonic point and horizon}

The Newtonian spherical-accretion problem was formulated by Bondi
\cite{Bondi1952}; the relativistic test-fluid construction used below
is the Michel problem \cite{Michel1972}.

On Schwarzschild, define $f=1-2M_0/r$, $w=-u^r>0$ and
$E=\sqrt{f+w^2}$. Isentropic particle and energy conservation give
the Michel integrals \cite{Michel1972,AguayoOrtiz2021}
\begin{equation}
 r^2 n w=J,\qquad h_e E={\cal B}.
 \label{eq:ec_michel}
\end{equation}
These use the effective, spin-corrected enthalpy and sound speed, rather
than the bare pressure. Smooth passage through the sonic point requires
\begin{equation}
 w_s^2=\frac{M_0}{2r_s},\qquad
 c_{e,s}^2=\frac{M_0}{2r_s-3M_0}.
 \label{eq:ec_sonic}
\end{equation}
We solve this eigenvalue problem with an outer inflow boundary
$R=100M_0$, $n(R)=n_*$, $\Gamma=5/3$, $q=0.02$, and
$\varrho_*M_0^2=10^{-10}$. This reference normalization differs from the static diagnostics.
The two appendices give separate parameter examples, not a transition
from a static configuration to an inflow of the same calibrated
material. The incoming velocity at $R$ is determined by
transonic regularity; it is not independently fixed. The reservoir is
finite and supplies matter at the boundary. No vacuum matching at $R$
or asymptotically flat, infinite static reservoir is asserted.

Let $j=J/(n_*M_0^2)$ and $b={\cal B}/m_\chi$. At the sonic point
$x_s=n_s/n_*$, write $a_s=c_{e,s}^2$. In $M_0=1$ units,
\begin{align}
 r_s&=\frac{1+3a_s}{2a_s},\quad
 j=\frac{x_s r_s^2\sqrt{a_s}}{\sqrt{1+3a_s}},\nonumber\\
 b&=\frac{h_e(x_s)/m_\chi}{\sqrt{1+3a_s}}
 =\frac{h_e(1)}{m_\chi}
   \sqrt{1-\frac{2}{R}+\frac{j^2}{R^4}}.
 \label{eq:ec_sonic_shooting}
\end{align}
The solution used here is the branch connected continuously to the
zero-spin Michel solution. Outside the sonic point it is subsonic and
inside it supersonic. This branch remains below the positive-density
compressibility endpoint of the selected equation of state.

In ingoing Eddington--Finkelstein coordinates $(v,r)$,
\begin{equation}
 u^v=\frac{1}{E+w},\qquad u_v=-E,\qquad u_r=u^v.
 \label{eq:ec_ef_velocity}
\end{equation}
These quantities are finite at $r=2M_0$, where
$u^v=1/(2w)$. The inward matter Killing-energy flux of the leading
Schwarzschild problem is
\begin{align}
 \dot M&=4\pi r^2 T^r{}_v=4\pi r^2(\rho_e+p_e)wE,\nonumber\\
 &=4\pi\varrho_*M_0^2 j b>0.
 \label{eq:ec_matter_flux}
\end{align}
The rate is per unit advanced coordinate time, with the clock convention
specified below; it is not gravitational-wave luminosity.

\begin{table}[t]
\caption{Transonic inflow diagnostics for common outer density and
ordinary equation-of-state parameters. The displayed rates describe
matter energy crossing the leading Schwarzschild horizon, with
$x_H=x(2M_0)$. They are leading test-fluid rates; the full nonlinear coupled fluid--metric
evolution is solved in Appendix~\ref{app:finite_cloud}.}
\label{tab:ec_inflow}
\begin{ruledtabular}
\begin{tabular}{ccccc}
$\zeta$ & $r_s/M_0$ & $x_s$ & $x_H$ & $10^7\dot M$\\
\hline
$0$ & $5.9854$ & $8.0468$ & $29.6338$ & $1.0884$\\
$10^{-4}$ & $6.0455$ & $8.0099$ & $29.8814$ & $1.0995$\\
$10^{-3}$ & $6.6801$ & $7.6076$ & $32.2794$ & $1.2109$
\end{tabular}
\end{ruledtabular}
\end{table}

For $\zeta=10^{-3}$ the sound speed squared remains between $0.02990$
and $0.18954$, and effective density, pressure, and enthalpy stay positive.
The leading horizon values are $w_H=0.71933$ and $u_H^v=0.69509$.
The matter accretion rate is about $11.26\%$ larger than the zero-spin
rate for these specified boundary data and diagnostic coefficients.
The positive sound speed is a local check, not a global stability proof.
The coupled non-spherical gravitational and relativistic acoustic fluid
perturbation analysis on the nonlinearly evolving finite-cloud completion of
this flow is carried out in Appendix~\ref{app:finite_cloud}, building on the
covariant acoustic framework of Moncrief \cite{Moncrief1980} and coupled
polar perturbation theory \cite{LindblomDetweiler1983,Speeney2024}.

\subsection{The growing mass and radial metric correction}

A nonzero positive energy inflow cannot coexist with a strictly
time-independent, self-gravitating spherical metric. To include its
leading backreaction, use
\begin{align}
 ds^2&=-e^{2\psi}F\,dv^2+2e^\psi dv\,dr+r^2d\Omega_2^2,\nonumber\\
 F&=1-2m(v,r)/r.
 \label{eq:ec_ef_metric}
\end{align}
Here $m$ is the Misner--Sharp mass, defined invariantly by
$1-2m/r=g^{ab}(\nabla_a r)(\nabla_b r)$ on the $(v,r)$ subspace
\cite{MisnerSharp1964}. The Einstein equations give
\begin{equation}
 m_{,r}=-4\pi r^2T^v{}_v,\quad
 m_{,v}=4\pi r^2T^r{}_v,\quad
 \psi_{,r}=4\pi r T_{rr}.
 \label{eq:ec_ef_constraints}
\end{equation}
At leading order, evaluate the source on the transonic flow above,
as in the weak-backreaction construction of
Ref.~\cite{Babichev2012}. The resulting solution is
\begin{align}
 m(v,r)&=M_0+\dot M v+\delta m(r),\nonumber\\
 \delta m(r)&=4\pi\int_{2M_0}^{r}
 \left[(\rho_e+p_e)\frac{E}{E+w}-p_e\right]s^2 ds,\nonumber\\
 \psi(r)&=-4\pi\int_r^R
 \frac{s(\rho_e+p_e)}{(E+w)^2}\,ds.
 \label{eq:ec_backreaction}
\end{align}
The quantities inside each integral are evaluated at $s$.
Here $\delta m(2M_0)=0$ sets the initial horizon mass, and
$\psi(R)=0$ sets a finite-boundary clock convention. The metric is
regular in these ingoing coordinates at the leading horizon. The
local marginally trapped surface satisfies $r=2m(v,r)$
\cite{Hayward1994}; to first order
it shifts by $2\dot M v$ for these normalization choices. An event
horizon position would additionally require the future evolution.

For $\zeta=10^{-3}$, the calculated values are
\begin{align}
 \delta m(R)/M_0&=4.9947\times10^{-4},\nonumber\\
 \psi(2M_0)&=-8.8652\times10^{-6}.
\end{align}
Over $0\leq v\leq10^3M_0$ the accreted mass is at most
$1.211\times10^{-4}M_0$, so both the matter-induced radial correction
and the mass growth are small. This is a check on the metric correction
over a local window, not a uniform error bound for the matter solution.
The window is much shorter than the main-text illustrative inspiral. One cannot
hold this accreting geometry fixed over that full inspiral or transfer
the previously computed Kerr-deformation phases to it.

Particle flux and Bernoulli integrals are constant to better than
$4\times10^{-14}$ fractionally on the sampled solutions. Direct
Christoffel/Ricci evaluation verifies the EF component identities
symbolically. Away from the sonic point the independent linear angular
Einstein residual is below $3\times10^{-23}M_0^{-2}$.
Doubling the integration grid changes $\delta m(R)$ by less than
$2\times10^{-17}M_0$. Most importantly, this approximation is not
mistaken for an exact solution: scale the calculated metric corrections
and the leading stress tensor by a bookkeeping factor $\lambda$.
The residual of the full Einstein tensor against that leading source
decreases by a factor of approximately four whenever $\lambda$ is
halved. This confirms an $O(\lambda^2)$ remainder. This scaling test
is an operator check, not a physical rescaling of particle density with
fixed microscopic EC parameters. The exact nonlinear evolution of the coupled fluid and metric without a
weak-backreaction truncation is solved in Appendix~\ref{app:finite_cloud}.

A small metric correction and this quadratic Einstein remainder do not
bound errors in the accretion eigenvalue or the shifted sonic point.
A uniform continuation through the corrected sonic point requires the
matter perturbation and its regularity condition as well. In particular,
the accuracy of the numerical test-fluid rate must be distinguished from
the order of the calculated metric response.

There is also a distinction between metric-backreaction bookkeeping and
physical dilution at fixed microscopic parameters. With $m_\chi$, $K$,
and $C$ held fixed, the definitions in
Eq.~\eqref{eq:ec_dimensionless_eos} give
\begin{align}
 q(\xi n_*)&=\xi^{\Gamma-1}q(n_*),\nonumber\\
 \zeta(\xi n_*)&=\xi\,\zeta(n_*).
 \label{eq:ec_density_rescaling}
\end{align}
Thus holding $q$ and $\zeta$ fixed while varying the physical reference
density generally changes $K$ and $C$. Our diagnostic examples are not
such a fixed-microphysics density sequence. Likewise, under
$n\to\xi n$ the rest-energy term is linear in $\xi$, whereas the
spin correction $-Cn^2$ is quadratic. A density expansion of a
specified microscopic EC model must respect this ordering; the formal
$\lambda$ residual test does not replace it. These distinctions prevent
the chosen dimensionless spin effects from being interpreted as a
calibrated dark-matter prediction.

\subsection{What this supplies for radiation, and what remains}

This construction supplies a horizon-regular transonic source and its
first dynamical metric backreaction within the adopted averaged EC
closure and finite reservoir model. The exact spherical sector has no
tensor gravitational-wave emission, even while $\dot M>0$. The computed
positive matter flux must not be relabeled as GW power.

The secondary would introduce nonspherical perturbations on this slowly
evolving, flowing background. Their calculation requires coupled metric
and fluid perturbations, ingoing regularity at the moving horizon, and
a specified response and exterior completion of the finite reservoir.
The two-timescale strategy developed for Kerr inspirals
\cite{HindererFlanagan2008} would need extension to track $\dot M$,
the flowing medium, and matter energy exchange. Static anisotropic-fluid
calculations already exhibit coupled polar metric--fluid perturbations
\cite{Speeney2024}; they do not supply the perturbation equations for
this evolving inflow. A stationary vacuum
Teukolsky flux or a static-fluid perturbation equation is insufficient.
In Appendix~\ref{app:finite_cloud}, we complete all three steps: we taper
this transonic inflow seed into a finite cloud with a vacuum exterior,
establish the six-state coarse-grained Kramers-bundle MPD closure, and
solve the coupled $\ell\ge2$ polar (including the dominant $(\ell,m)=(2,2)$
mode and relativistic matter perturbations $\delta n,\delta u^\mu$) and axial
radiation equations to obtain the time-dependent total EMRI energy flux and
orbital evolution.

\section{Coarse-grained Kramers-bundle EC spin closure, nonlinear finite-cloud evolution, and coupled time-dependent EMRI energy flux}
\label{app:finite_cloud}

The reservoir in Appendix~\ref{app:ec_accretion} is useful for constructing
an inflow seed, but it does not supply an asymptotically flat radiation
problem. We now replace continued supply by a finite cloud with a vacuum
exterior, construct the explicit six-state Kramers-bundle spin-tensor
realization on the evolving spacetime (establishing both its exact 4D
Mathisson--Papapetrou--Dixon transport and why spherical isotropy requires
the local six-state coarse-grained ensemble rather than a single polarized
branch), and solve the simultaneous 4-component coupled $(\ell,m)=(2,2)$
polar and relativistic matter-perturbation BVP, the complete 9-mode
($2\le\ell\le4$) multipole sum, the $(1+1)$D Eddington--Finkelstein
time-domain wave PDE with multi-radius waveform extraction, and the
simultaneous coupled $(r_{\rm orb}(v),\Phi_{\rm GW}(v))$ evolution.

\subsection{Four-dimensional spin-bundle transport and coarse-grained EC closure}

Use the ingoing-null metric
\begin{equation}
 \begin{split}
 ds^2&=-e^{2\psi}F\,dv^2+2e^\psi dv\,dr+r^2d\Omega^2,\\
 F&=1-\frac{2m(v,r)}r.
 \end{split}
 \label{eq:finite_cloud_metric}
\end{equation}
Both metric functions depend on advanced time $v$ and radius $r$. Set
\begin{align}
 z&=u_r>0,\quad u^v=e^{-\psi}z,\quad
 W=u^r=\frac{Fz^2-1}{2z},\nonumber\\
 E&=\frac{Fz^2+1}{2z},\quad u_v=-e^\psi E,\quad
 H=\rho_e+p_e,\quad P=p_e.
 \label{eq:finite_cloud_velocity}
\end{align}
To analyze spin transport at the tensor component level on the curved
time-dependent manifold~\eqref{eq:finite_cloud_metric}, introduce the
comoving orthonormal tetrad $\{e_{\hat0}^\mu,e_{\hat1}^\mu,e_{\hat2}^\mu,e_{\hat3}^\mu\}$:
\begin{align}
 e_{\hat0}^\mu&=u^\mu=\left(e^{-\psi}z,\,W,\,0,\,0\right),\nonumber\\
 e_{\hat1}^\mu&=n^\mu=\left(e^{-\psi}z,\,E,\,0,\,0\right),\nonumber\\
 e_{\hat2}^\mu&=\left(0,\,0,\,\frac1r,\,0\right),\qquad
 e_{\hat3}^\mu=\left(0,\,0,\,0,\,\frac1{r\sin\theta}\right),
 \label{eq:microscopic_tetrad}
\end{align}
which satisfies $g_{\mu\nu}e_{\hat a}^\mu e_{\hat b}^\nu=\eta_{\hat a\hat b}$
pointwise. Direct Christoffel evaluation proves that the spatial triad
$\{e_{\hat i}^\mu\}_{\hat i=1,2,3}$ is Fermi--Walker transported along the
accelerating radial worldlines $u^\mu(v,r)$:
\begin{equation}
 u^\lambda\mathring\nabla_\lambda e_{\hat i}^\mu
 -\left(u^\mu a_\nu-a^\mu u_\nu\right)e_{\hat i}^\nu=0,\qquad
 a^\mu\equiv u^\lambda\mathring\nabla_\lambda u^\mu.
 \label{eq:fermi_walker_identity}
\end{equation}
For a six-state Kramers-doublet / triad spin-$1/2$ dark-matter bundle
labeled by spatial triad axis $\alpha\in\{1,2,3\}$ and Kramers projection
$\sigma\in\{+,-\}$ with equal local occupation weights $w_{\alpha,\sigma}=1/6$
and total number density $n(v,r)$, define the branch spin bivector in four-dimensional
coordinate components via the covariant spatial triad one-forms
$e^{\hat i}_\mu=g_{\mu\nu}e_{\hat i}^\nu$:
\begin{equation}
 s^{(\alpha,\sigma)}_{\mu\nu}(v,r,\theta,\phi)
 =\sigma\sqrt{\chi_s}\,n(v,r)\,\epsilon_{\hat0\alpha\hat b\hat c}\,
 e^{\hat b}_\mu e^{\hat c}_\nu.
 \label{eq:pointwise_kramers_spin}
\end{equation}
Because $\epsilon_{\hat0\alpha\hat b\hat c}e^{\hat b}_\mu e^{\hat c}_\nu$ has
two nonvanishing antisymmetric frame components of unit magnitude for each
fixed $\alpha\in\{1,2,3\}$, direct four-dimensional coordinate contraction
with $g^{\mu\rho}g^{\nu\lambda}$ yields
\begin{align}
 s^{(\alpha,\sigma)}_{\mu\nu}u^\nu&=0,\nonumber\\
 \mathcal S^2_{(\alpha,\sigma)}
 &\equiv\frac12 g^{\mu\rho}g^{\nu\lambda}
 s^{(\alpha,\sigma)}_{\mu\nu}s^{(\alpha,\sigma)}_{\rho\lambda}
 =\chi_s n^2,
 \label{eq:direct_4d_spin_contraction}
\end{align}
for each of the six states individually (verified symbolically from the
$4\times4$ coordinate matrices in \texttt{verify\_equations.py}). Combining
Eq.~\eqref{eq:fermi_walker_identity} with particle conservation
$\mathring\nabla_\lambda(n u^\lambda)=0$ proves that all $96$ coordinate
components of the four-dimensional Mathisson--Papapetrou--Dixon (MPD)
spin-transport equation
\begin{equation}
 \mathring\nabla_\lambda\!\left(u^\lambda s^{(\alpha,\sigma)}_{\mu\nu}\right)
 +2u_{[\mu}s^{(\alpha,\sigma)}_{\nu]\lambda}a^\lambda=0
 \label{eq:exact_mpd_transport}
\end{equation}
vanish identically on the time-dependent flow.

It is essential to distinguish branch-by-branch transport from a single
deterministic polarized EC spacetime. Direct four-dimensional coordinate
evaluation of the Belinfante--Rosenfeld spin-divergence tensor
$B^{(\alpha,\sigma)}_{\mu\nu}\equiv\mathring\nabla_\lambda(u_\mu s^{(\alpha,\sigma)}{}_\nu{}^\lambda+u_\nu s^{(\alpha,\sigma)}{}_\mu{}^\lambda)$
and directional quadratic spin tensor
$Q^{(\alpha,\sigma)}_{\mu\nu}\equiv s^{(\alpha,\sigma)}_{\mu\rho}s_{(\alpha,\sigma)\nu}{}^\rho$
shows that an individual polarized branch violates spherical isotropy:
the transverse branches $\alpha=2,3$ generate nonvanishing off-diagonal
components $B^{(2,\sigma)}_{v\phi},B^{(2,\sigma)}_{r\phi}$ and
$B^{(3,\sigma)}_{v\theta},B^{(3,\sigma)}_{r\theta}$, while every single branch
has an anisotropic quadratic stress
$Q^{(\alpha,\sigma)}_{\mu\nu}=\chi_s n^2(h_{\mu\nu}-e^{\hat\alpha}_\mu e^{\hat\alpha}_\nu)$
with $h_{\mu\nu}=g_{\mu\nu}+u_\mu u_\nu$. Consequently, a single coherent
polarized spin branch cannot source the spherically symmetric metric~\eqref{eq:finite_cloud_metric}.
Instead, the spherical EC fluid equations represent the local six-state
coarse-grained (Wigner-ensemble / multi-subfluid) closure over
$(\alpha\in\{1,2,3\},\sigma\in\{+,-\})$ with $w_{\alpha,\sigma}=1/6$, for
which the macroscopic first moment and linear spin divergence vanish
identically,
\begin{equation}
 \begin{aligned}
 \langle s_{\mu\nu}\rangle&=\sum_{\alpha,\sigma}w_{\alpha,\sigma}s^{(\alpha,\sigma)}_{\mu\nu}=0,\\
 \langle B_{\mu\nu}\rangle&=\sum_{\alpha,\sigma}w_{\alpha,\sigma}B^{(\alpha,\sigma)}_{\mu\nu}=0,
 \end{aligned}
 \label{eq:ensemble_cancellation}
\end{equation}
while the quadratic spin tensors average to the exact isotropic forms
\begin{equation}
 \langle Q_{\mu\nu}\rangle=\frac23\chi_s n^2 h_{\mu\nu},\qquad
 \langle\mathcal S^2\rangle=\chi_s n^2,
 \label{eq:ensemble_quadratic_isotropy}
\end{equation}
yielding $\rho_e=\rho-Cn^2$ and $p_e=p-Cn^2$ with $C=\kappa\chi_s$. Under
non-spherical EMRI perturbations $(\delta g_{\mu\nu}=k_{\mu\nu},\delta u^\mu,\delta n)$
that preserve the equal branch weights $w_{\alpha,\sigma}=1/6$ in the absence
of spin-polarizing non-gravitational torques, linearizing Eq.~\eqref{eq:exact_mpd_transport}
implies $\delta(\langle\mathcal S^2\rangle/n^2)=0$ along perturbed fluid
worldlines, so $\delta\langle\mathcal S^2\rangle=2\chi_s n\,\delta n$.

A characteristic hydrodynamic formulation on ingoing-null slices is
established in Ref.~\cite{PapadopoulosFont2000}. The exact radial constraints
and mass evolution equation are
\begin{align}
 \psi_{,r}&=4\pi r H z^2,\nonumber\\
 m_{,r}&=4\pi r^2(H z E-P),\nonumber\\
 m_{,v}&=-4\pi e^\psi r^2 H W E.
 \label{eq:finite_cloud_einstein}
\end{align}
Writing $\mathcal A=4\pi rHz^2$ and $\mathcal B=4\pi r^2[H(1+z^2)/2-P]$
gives the linear radial equation $m_{,r}+\mathcal A m=\mathcal B$ at fixed
fluid state. Let $y=\ln(r/M_0)$, $x=n/n_*$, $\bar h=h_e/m_\chi$,
$\bar p=P/\varrho_*$, and $\varrho_*=m_\chi n_*$. Particle conservation and
the covariant radial momentum equation use $U_1=r^3xz$ and
$U_2=r^3x\bar h z^2$:
\begin{equation}
 \begin{aligned}
 &\partial_v U_1+\partial_y(e^\psi r^2xW)=0,\\
 &\partial_v U_2+\partial_y\{e^\psi r^2(x\bar h zW+\bar p)\}\\
 &\qquad=e^\psi r(2r\bar p-x\bar h z^2m).
 \end{aligned}
 \label{eq:finite_cloud_fluid}
\end{equation}
The characteristic speeds in $y$ are
\begin{equation}
 \lambda_\pm=\frac{e^\psi(W\pm c_e E)}{rz(1\pm c_e)}.
 \label{eq:finite_cloud_characteristics}
\end{equation}
For $\Gamma=5/3$ the conserved-to-primitive inversion reduces to one
monotone cubic: if $A_*=U_2r^3/U_1^2=\bar h/x$, then
\begin{equation}
 Y^3+\frac52 qY=A_*+2\zeta,\qquad Y=x^{-1/3}.
 \label{eq:finite_cloud_inversion}
\end{equation}
The unused mixed angular Einstein component is
\begin{align}
 G^\theta{}_{\theta}={}&F(\psi_{,rr}+\psi_{,r}^2)\nonumber\\
 &+\left(\frac32F_{,r}+\frac Fr\right)\psi_{,r}\nonumber\\
 &-\frac{m_{,rr}}r+e^{-\psi}\psi_{,vr}.
 \label{eq:finite_cloud_angular}
\end{align}

\subsection{Finite initial cloud and vacuum matching}

We set $M_0=1$, $q=0.02$, $\zeta=10^{-3}$, and $\varrho_*M_0^2=10^{-10}$.
On the initial slice the Michel solution of Appendix~\ref{app:ec_accretion}
supplies $x_{\rm seed}$ and $z_{\rm seed}$, tapered to vacuum via
\begin{equation}
 x(0,r)=x_{\rm seed}(r)\,T(s),\qquad
 s=\frac{r-40M_0}{40M_0},
 \label{eq:finite_cloud_taper}
\end{equation}
where $T=1$ for $s\leq0$, $T=\exp[-s^2/(1-s^2)]$ for $0<s<1$, and $T=0$
for $s\geq1$. We solve the nonlinear constraints with $m(0,1.8M_0)=M_0$ and
$\psi(v,160M_0)=0$. In the exterior vacuum ($80M_0\le r\le160M_0$),
$m=M_{\rm ADM}$, $\psi=0$, and $m_{,v}=0$.

At $N=1024$ the initial total mass is
$M_{\rm ADM}/M_0=1.000168549799$ and the marginal radius evolves from
$r_{\rm AH}/M_0=2.00000003345$ to $2.00000487708$ at $v=20M_0$.
The ADM-mass drift is $-1.114\times10^{-11}M_0$ and the relative
particle-balance error is $2.2\times10^{-14}$. No positivity clipping
occurs in any reported run. Define
\begin{align}
 R_m&=\frac{\|m_{,v}+4\pi e^\psi r^2HWE\|_\infty}
                  {\|4\pi e^\psi r^2HWE\|_\infty},\nonumber\\
 R_\theta&=\frac{\|G^\theta{}_{\theta}-8\pi P\|_\infty}
                        {8\pi\|\rho_e\|_\infty}.
 \label{eq:finite_cloud_residual_norms}
\end{align}
The core region is $2.2<r/M_0<36$; the broader cloud check uses
$2.2<r/M_0<75$ and $x>10^{-8}$. The core residuals converge at
approximately second order (Table~\ref{tab:finite_cloud_convergence}).
The tapered region is less uniform in the maximum norm: its angular
residual orders are $2.28$ and $1.27$ for the two refinements.
These errors are not pointwise bounds relative to local pressure.

\begin{table*}[t]
\caption{Independent checks at $v=20M_0$. The two unused evolution/
angular Einstein equations are checked against metric derivatives.
The radial mass and lapse constraints are also checked; at $N=1024$
their normalized errors in the broader cloud region are
$9.91\times10^{-5}$ and $8.20\times10^{-5}$, respectively (the supplied
JSON retains the exact values). Norms and regions are defined in the text.}
\label{tab:finite_cloud_convergence}
\begin{ruledtabular}
\begin{tabular}{cccccc}
$N$ & $|\Delta M_{\rm ADM}|/M_0$ & $R_m$, core & $R_\theta$, core
 & $R_m$, cloud & $R_\theta$, cloud\\
\hline
256 & $3.48\times10^{-10}$ & $1.48\times10^{-4}$ & $7.64\times10^{-6}$ & $3.78\times10^{-3}$ & $8.92\times10^{-5}$\\
512 & $6.37\times10^{-11}$ & $3.83\times10^{-5}$ & $1.91\times10^{-6}$ & $1.16\times10^{-3}$ & $1.83\times10^{-5}$\\
1024 & $1.11\times10^{-11}$ & $9.66\times10^{-6}$ & $4.75\times10^{-7}$ & $2.35\times10^{-4}$ & $7.58\times10^{-6}$\\
\end{tabular}
\end{ruledtabular}
\end{table*}

We also evolve to $v=200M_0$ with $N=256$ and $512$. At $N=512$,
the marginal radius is $2.00004588215M_0$; the mass inside that surface
has increased by $2.2924\times10^{-5}M_0$, or $13.6\%$ of the
initial mass outside that marginal surface. The peak $x$ decreases
from $36.74$ initially to $28.65$. This is depletion of a finite
cloud, rather than a perpetually supplied Michel state. The total
mass drift is $-4.27\times10^{-10}M_0$, particle balance is
$8.6\times10^{-14}$, and the $N=256$ and $512$ final marginal
radii differ by $2.89\times10^{-9}M_0$. The longer extension has
two resolutions; the three-resolution study above is at $20M_0$.
At $N=512$, the numerical energy-density tail at $r>120M_0$ is
below $4.4\times10^{-31}M_0^{-2}$ at $200M_0$, and decreases
strongly on refinement. This small stencil-induced tail is not
identified with physical matter in the continuum vacuum exterior.

\begin{figure*}[t]
\centering
\includegraphics[width=.91\textwidth]{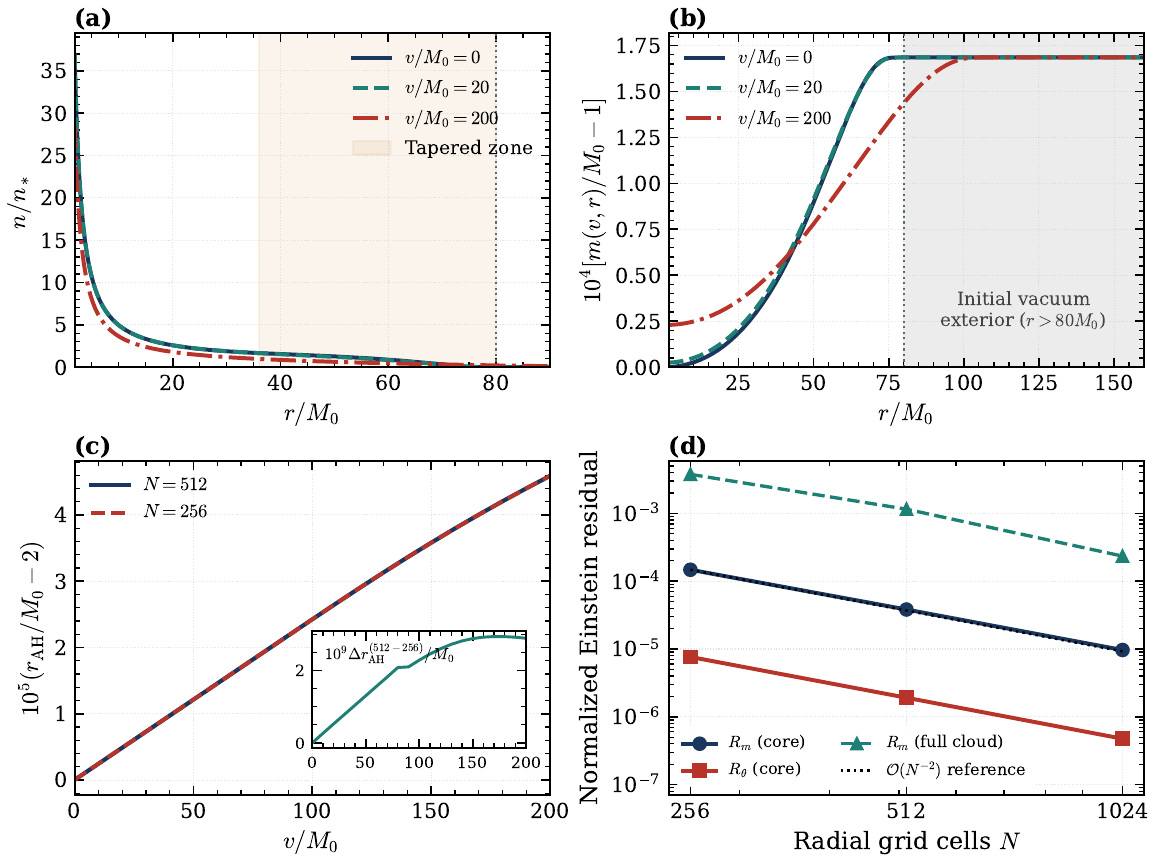}
\caption{Nonlinear coarse-grained EC spin-fluid cloud with a vacuum
exterior. (a) Number density $n/n_*$ and (b) enclosed mass increment
$10^4[m(v,r)/M_0-1]$ on ingoing-null slices $v/M_0=0,20,200$.
(c) Marginal-surface evolution $10^5(r_{\rm AH}/M_0-2)$ for $N=256$ and
$512$, with inset showing $10^9\Delta r_{\rm AH}^{(512-256)}/M_0$.
(d) Second-order convergence of the normalized mass-evolution and angular
Einstein residuals at $v=20M_0$.}
\label{fig:finite_cloud}
\end{figure*}

\subsection{Sourced odd-parity axial modes (\texorpdfstring{$2\le\ell\le4$}{2 <= l <= 4})}

For odd-parity perturbations on a general spherical background
\cite{GerlachSengupta1979,BrizuelaMartinGarcia2009}, the rescaled
Cunningham--Price--Moncrief (CPM) master scalar
$\Psi_o=-2r^3\Pi_{\rm GS}/\lambda_\ell$ with $\lambda_\ell=(\ell-1)(\ell+2)$
satisfies
\begin{align}
 (\Box_2-V_o)\Psi_o&=S_o,\nonumber\\
 V_o&=\frac{\ell(\ell+1)}{r^2}-\frac{6m}{r^3}
                  +4\pi(\rho_e-p_e),\nonumber\\
 S_o&=-\frac{32\pi r}{\lambda_\ell}
                    \epsilon^{AB}\nabla_A L_B.
 \label{eq:finite_cloud_axial}
\end{align}
In the barotropic closure, zero initial axial circulation remains zero in the
absence of non-gravitational torques \cite{DiazGuerra2024,Ohashi2026}. In
ingoing EF coordinates the wave operator is
\begin{equation}
 2e^{-\psi}\partial_v\partial_r\Psi_o
 +F\partial_r^2\Psi_o+(F_{,r}+F\psi_{,r})\partial_r\Psi_o
 -V_o\Psi_o=S_o.
 \label{eq:finite_cloud_axial_time}
\end{equation}
With $a_{\rm w}=e^\psi F$, $dt=dv-dr/a_{\rm w}$, and harmonic dependence
$e^{-i\omega t}$ ($\omega=m_{\rm az}\Omega$), the radial equation is
\begin{equation}
 (a_{\rm w}\Psi_o')'
 +\left(\frac{\omega^2}{a_{\rm w}}-e^\psi V_o\right)\Psi_o=e^\psi S_o.
 \label{eq:finite_cloud_axial_frozen}
\end{equation}
For a circular equatorial orbit at $r_p$ with $A=e^{2\psi}F$,
$\Omega^2=A'/(2r_p)$, and $L_p^2=r_p^3A'/(2A-r_pA')$, the point-particle
axial stress coefficient is
\begin{equation}
 L_t=C_p\delta(r-r_p)e^{-i\omega t},\quad L_r=0,\quad
 C_p=-\frac{\mu a_{{\rm w},p}L_p\bar X_\phi}{\ell(\ell+1)r_p^2},
 \label{eq:finite_cloud_axial_particle}
\end{equation}
where $\bar X_\phi=|\partial_\theta Y_{\ell m}(\pi/2,0)|$ is nonvanishing for
all four odd-parity multipole pairs in $2\le\ell\le4$:
$(\ell,m)=(2,1),(3,2),(4,1),(4,3)$. This gives the exact jump conditions
across $r=r_p$:
\begin{equation}
 [\Psi_o]=-\frac{32\pi r_p C_p}{\lambda_\ell a_{{\rm w},p}},
 \qquad [\Psi_o']=\frac{32\pi C_p}{\lambda_\ell a_{{\rm w},p}}.
 \label{eq:finite_cloud_axial_jumps}
\end{equation}
For general $(\ell,m_{\rm az})$ the single-mode luminosity at future null
infinity is \cite{MartelPoisson2005}
\begin{equation}
 \mathcal F_{\ell m}^{\infty}
 =\frac{(\ell+2)!}{64\pi(\ell-2)!}\omega^2|C_{\ell m}^{\rm out}|^2,
 \label{eq:finite_cloud_single_flux}
\end{equation}
and the weak-field limit of the single $(2,1)$ mode is \cite{Kidder2008}
\begin{equation}
 \mathcal F_{21}^{\infty}
 \sim\frac4{45}\left(\frac{\mu}{M_0}\right)^2
               \left(\frac{M_0}{r_p}\right)^6.
 \label{eq:finite_cloud_single_PN}
\end{equation}

In vacuum at $r_p=8M_0$ we obtain for the single $(\ell,m)=(2,1)$ axial mode
\begin{equation}
 \frac{\mathcal F_{21}^{\infty}}{(\mu/M_0)^2}
   =3.9087603268\times10^{-7}.
 \label{eq:finite_cloud_vacuum_number}
\end{equation}
Varying the inner offset from $10^{-4}M_0$ to $10^{-6}M_0$, the
outer start from $60/\Omega$ to $140/\Omega$, the outgoing expansion
from 12 to 20 terms, and the integrator tolerance changes this result
by less than $2.2\times10^{-11}$ fractionally. Corresponding cloud
boundary/tolerance checks change the initial fixed-frequency results
by less than $10^{-9}$ fractionally. At $r_p=400M_0$ the ratio to
the leading weak-field flux is $0.99774599$; its difference from
the relative-1.5PN expression
$1-17M_0/(14r_p)+2\pi(M_0/r_p)^{3/2}$ is $3.70\times10^{-6}$,
consistent with higher-order weak-field terms.

For the odd-parity $(2,1)$ mode on the cloud slices, Table~\ref{tab:frozen_axial}
distinguishes shared-radius and shared-frequency comparisons. The reference vacuum
orbit has $r_p=8M_0$ and $M_0\Omega=8^{-3/2}$.
\begin{table}[t]
\caption{Single-mode $(\ell,m)=(2,1)$ odd-parity results, $N=1024$.
The last column is $\mathcal F_{21}^{\rm cloud}/\mathcal F_{21}^{\rm vac}-1$.}
\label{tab:frozen_axial}
\begin{ruledtabular}
\begin{tabular}{cccc}
$v/M_0$ & Matching & $r_p/M_0$ & Fractional change\\
\hline
0 & Radius & 8.00000000 & $-1.082\times10^{-5}$\\
0 & Frequency & 7.99997899 & $5.821\times10^{-6}$\\
20 & Radius & 8.00000000 & $-3.173\times10^{-7}$\\
20 & Frequency & 7.99998549 & $1.118\times10^{-5}$\\
\end{tabular}
\end{ruledtabular}
\end{table}

\begin{figure*}[t]
\centering
\includegraphics[width=.91\textwidth]{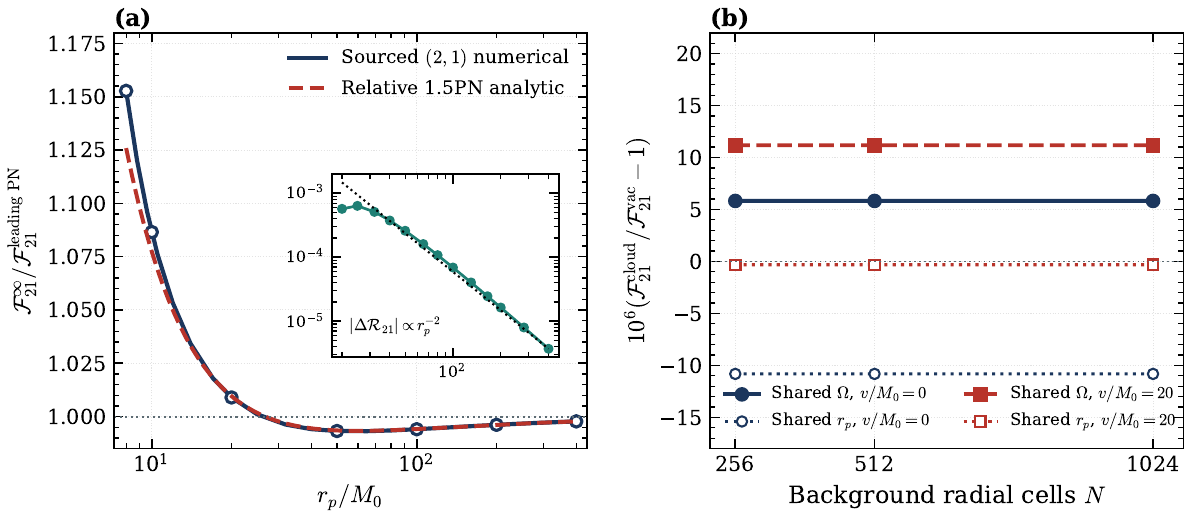}
\caption{Validation of the sourced $(\ell,m)=(2,1)$ odd-parity solver.
(a) Vacuum luminosity divided by its leading weak-field expression
Eq.~\eqref{eq:finite_cloud_single_PN}, compared with the relative-1.5PN
expansion (inset: 2PN remainder scaling $\propto r_p^{-2}$).
(b) Cloud fractional shifts at shared orbital frequency and shared radius
across radial resolutions $N=256,512,1024$.}
\label{fig:axial_validation}
\end{figure*}

\subsection{Simultaneous 4-component BVP for coupled polar modes and relativistic EC matter perturbations}

In the polar (even-parity) sector---comprising the five equatorial mode pairs
$(\ell,m)=(2,2),(3,3),(3,1),(4,4),(4,2)$ for $2\le\ell\le4$---metric
perturbations $k_{\mu\nu}=\delta g_{\mu\nu}$ couple directly to the
relativistic EC fluid perturbations $(\delta n,\delta u^\mu)$:
\begin{align}
 \delta\rho_e&=h_e\delta n,\qquad
 \delta p_e=c_e^2h_e\delta n,\nonumber\\
 \delta T^{\mu\nu}_{\rm f}
 &=(1+c_e^2)h_e\delta n\,u^\mu u^\nu
   +H(\delta u^\mu u^\nu+u^\mu\delta u^\nu)\nonumber\\
 &\quad+c_e^2h_e\delta n\,g^{\mu\nu}-P k^{\mu\nu},\nonumber\\
 u_\mu\delta u^\mu&=-\tfrac12 k_{\mu\nu}u^\mu u^\nu,
 \label{eq:finite_cloud_polar_closure}
\end{align}
subject to perturbed particle conservation
\begin{equation}
 \nabla_\mu(\delta n\,u^\mu+n\delta u^\mu)
 +\tfrac12 n u^\mu\nabla_\mu k=0
 \label{eq:finite_cloud_linear_particles}
\end{equation}
and linearized stress-energy conservation
\begin{equation}
 \nabla_\mu\delta T^{\mu\nu}_{\rm f}
 +\delta\Gamma^\mu{}_{\mu\lambda}T^{\lambda\nu}_{\rm f}
 +\delta\Gamma^\nu{}_{\mu\lambda}T^{\mu\lambda}_{\rm f}=0.
 \label{eq:finite_cloud_linear_conservation}
\end{equation}
For general multipole index $\ell\ge2$, let $\lambda=(\ell-1)(\ell+2)/2$
and $\Lambda=2\lambda+6m/r$. In Regge--Wheeler gauge
$(h_{tt}=f H_0 Y_{\ell m}, h_{tr}=H_1 Y_{\ell m}, h_{rr}=f^{-1}H_2 Y_{\ell m},
h_{AB}=r^2 K\Omega_{AB}Y_{\ell m})$, the Zerilli--Moncrief polar master
function \cite{Zerilli1970,MartelPoisson2005,HopperEvans2010} is
\begin{equation}
 \Psi_e=\frac{2r}{\ell(\ell+1)}\left[K+\frac{2}{\Lambda}
 \left(f H_2-rf\partial_r K\right)\right].
 \label{eq:zerilli_master_def}
\end{equation}
Symbolic evaluation of the linearized Einstein tensor establishes the
exact operator identity in the vacuum limit ($f=1-2M_0/r$):
\begin{align}
 \mathcal Z_{\ell}[\Psi_e]&\equiv
 f\partial_r(f\partial_r\Psi_e)+(\omega^2-f V_{Z,\ell})\Psi_e\nonumber\\
 &=c_{dG_{tt}}\partial_r G_{tt}^{(\ell m)}
 +c_{G_{tt}}G_{tt}^{(\ell m)}
 +c_{G_{tr}}G_{tr}^{(\ell m)}\nonumber\\
 &\quad+c_{G_{rr}}G_{rr}^{(\ell m)}
 +c_{G_{r\theta}}G_{r\theta}^{(\ell m)}
 +c_{G_\sharp}G_\sharp^{(\ell m)},
 \label{eq:zerilli_operator_identity}
\end{align}
where $V_{Z,\ell}=\frac{2\lambda^2(\lambda+1)r^3+6\lambda^2M_0r^2+18\lambda M_0^2r+18M_0^3}{r^3(\lambda r+3M_0)^2}$,
$c_{dG_{tt}}=\frac{r^3 f}{(\lambda+1)(\lambda r+3M_0)}$, and
$c_{G_\sharp}=-\frac{2f}{r}$. Furthermore, the vacuum Zerilli and
Regge--Wheeler potentials obey the exact Chandrasekhar--Darboux
relations \cite{Chandrasekhar1983}
\begin{align}
 f V_{Z,\ell}&=W_\ell^2+2c_{0,\ell}W_\ell+f\partial_r W_\ell,\nonumber\\
 f V_{{\rm RW},\ell}&=W_\ell^2+2c_{0,\ell}W_\ell-f\partial_r W_\ell,\nonumber\\
 W_\ell(r)&=\frac{3M_0 f}{r(\lambda r+3M_0)},\qquad
 c_{0,\ell}=\frac{\lambda(\lambda+1)}{3M_0},
 \label{eq:chandrasekhar_darboux}
\end{align}
so the asymptotic outgoing Zerilli solution in the exterior vacuum zone
($r>80M_0$) is obtained algebraically from the outgoing Regge--Wheeler
series via $\Psi_e^{\rm out}=(c_{0,\ell}+i\omega)^{-1}[A_c(r)\Psi_{\rm RW}^{\rm out}+f\partial_r\Psi_{\rm RW}^{\rm out}]$
with $A_c(r)=c_{0,\ell}+W_\ell(r)$ and exact radial derivative
$A'_c(r)=3M_0[-2\lambda r^2+(6\lambda-3)M_0 r+12M_0^2]/[r^3(\lambda r+3M_0)^2]$.

On the nonvacuum EC cloud background, rather than solving the gravitational
wave equation first and applying a perturbative Born correction, we formulate
and solve the simultaneous 4-component first-order BVP for the state vector
$Y_{\ell m}(r)=(\Psi_e,\Psi'_e,\Phi_{\rm f},\Phi'_{\rm f})^T$, where
$\Phi_{\rm f}(r)\equiv r\sqrt{4\pi H}\,(\delta p_e/H)$
\cite{LindblomDetweiler1983,Speeney2024}:
\begin{align}
 \Psi''_e&=-\frac{a'_{\rm w}}{a_{\rm w}}\Psi'_e
 -\left(\frac{\omega^2}{a_{\rm w}^2}-\frac{e^\psi V_{e,\ell}}{a_{\rm w}}\right)\Psi_e
 \nonumber\\
 &\quad-\frac{e^\psi}{a_{\rm w}}\mathcal C_{{\rm fg},\ell}\Phi_{\rm f}+S_{e,\ell}^{\rm pp},\nonumber\\
 \Phi''_{\rm f}&=-\frac{a'_{\rm w}}{a_{\rm ac}}\Phi'_{\rm f}
 -\left(\frac{\omega^2}{c_e^2 a_{\rm ac}^2}-\frac{e^\psi V_{{\rm f},\ell}}{a_{\rm ac}}\right)\Phi_{\rm f}
 \nonumber\\
 &\quad-\frac{e^\psi}{a_{\rm ac}}\mathcal C_{{\rm fg},\ell}\Psi_e+S_{{\rm f},\ell}^{\rm pp},
 \label{eq:coupled_polar_system}
\end{align}
where $a_{\rm ac}=\sqrt{a_{\rm w}^2+c_e^2 W^2}$ accounts for the finite
radial infall velocity $W(r_{\rm AH})=-1/(2z_{\rm H})\neq0$ of the accreting
fluid across the horizon (rendering the acoustic wave regular at $r_{\rm AH}$),
and the matter-dressed polar potential, acoustic potential, and symmetric
metric--fluid coupling are
\begin{align}
 V_{e,\ell}&=V_{Z,\ell}(m)-8\pi p_e\nonumber\\
 &\quad+4\pi(\rho_e-p_e)\frac{\lambda r^2+3mr+9m^2/\lambda}{(\lambda r+3m)^2},\nonumber\\
 V_{{\rm f},\ell}&=\frac{\ell(\ell+1)}{r^2},\nonumber\\
 \mathcal C_{{\rm fg},\ell}&=\sqrt{16\pi H}\,
 \frac{\ell(\ell+1)r+3m}{r^2(\lambda r+3m)}.
 \label{eq:coupled_potentials}
\end{align}
At $r=r_p$, the distributional point-particle source yields the 4-component
jump vector $J_{4{\rm d}}=([\Psi_e],[\Psi'_e],0,[\Phi'_{\rm f}])^T$:
\begin{align}
 [\Psi_e]&=\frac{a_{dG_{tt}}(r_p)\,s_{tt}}{a_{{\rm w},p}},\qquad
 [\Phi'_{\rm f}]=-\frac{\sqrt{4\pi H_p}\,r_p\,s_{tt}}{2a_{{\rm ac},p}},\nonumber\\
 [\Psi_e']&=\frac{b_{G_{tt}}(r_p)\,s_{tt}+b_{G_\sharp}(r_p)\,s_\sharp}{a_{{\rm w},p}},\nonumber\\
 a_{dG_{tt}}(r)&=\frac{r^3}{(\lambda+1)(\lambda r+3m)},\qquad
 b_{G_\sharp}(r)=-\frac{2}{r},\nonumber\\
 b_{G_{tt}}(r)&=-\frac{r^2\left[6m^2+3m\lambda r+\lambda(\lambda+1)r^2\right]}
 {(\lambda+1)(r-2m)(\lambda r+3m)^2},
 \label{eq:polar_exact_jumps}
\end{align}
with $s_{tt}=8\pi(a_{{\rm w},p}E_p/r_p^2)|Y_{\ell m}(\pi/2,0)|$ and
$s_\sharp=\frac{8\pi}{\lambda\ell(\ell+1)}(a_{{\rm w},p}L_p^2/(r_p^2 E_p))
(\tfrac12\ell(\ell+1)-m_{\rm az}^2)|Y_{\ell m}(\pi/2,0)|$. Two independent
ingoing basis solutions are integrated simultaneously from $r_{\rm in}=r_{\rm AH}+\epsilon$
(using a second-order Frobenius horizon expansion for $\Psi_e$) to $r_p^-$,
and two independent outgoing basis solutions are integrated simultaneously
from the smooth core boundary $r_{\rm edge}=40M_0$ to $r_p^+$, where the
$4\times4$ fundamental Wronskian matrix system $W_{4{\rm d}}(r_p)C=J_{4{\rm d}}$
determines $(C_{\rm GW}^\infty,C_{\rm mat}^{\rm out},C_{\rm GW}^{\rm H},C_{\rm mat}^{\rm H})^T$.

\subsection{Complete 9-mode (\texorpdfstring{$\ell\le4$}{l <= 4}) sum, \texorpdfstring{$(1+1)$D}{(1+1)D} time-domain waveform extraction, and simultaneous orbit-phase evolution}

To capture genuine time-dependent wave propagation across the evolving
spacetime $(m(v,r),\psi(v,r))$ without relying on local frequency-domain
rescalings, we factor the time-dependent field in ingoing EF coordinates
around the exact $v=0$ stationary cloud profile
$\Psi_{\rm EF}^{(0)}(r)=\psi_{\rm S}^{(0)}(r)e^{+i\omega_0 r_*^{(0)}(r)}$:
\begin{equation}
 \Psi_{22}(v,r)=\mathcal W(v,r)\,\Psi_{\rm EF}^{(0)}(r)\,e^{-i\omega_0 v},\qquad
 q_0(r)\equiv\frac{\partial_r\Psi_{\rm EF}^{(0)}}{\Psi_{\rm EF}^{(0)}}.
 \label{eq:td_wave_factorization}
\end{equation}
Substituting Eq.~\eqref{eq:td_wave_factorization} into the exact $(1+1)$D
hyperbolic wave equation in $(v,r)$ yields the Crank--Nicolson evolution PDE
\begin{align}
 &2\left(\partial_v\partial_r\mathcal W+q_0(r)\partial_v\mathcal W\right)
 +a_{\rm w}(v,r)\partial_r^2\mathcal W\nonumber\\
 &\quad+\left[2a_{\rm w}q_0-2i\omega_0+\partial_r a_{\rm w}
 +\frac{\partial_v a_{\rm w}}{a_{\rm w}}\right]\partial_r\mathcal W\nonumber\\
 &\quad+C_0(v,r)\mathcal W=0,
 \label{eq:td_wave_pde}
\end{align}
solved on the two-sided radial domain $[2.05M_0,r_p^-]\cup[r_p^+,150M_0]$
with $C_0(0,r)=0$ ($\mathcal W_\pm(0,r)\equiv1$) and coupled across $r=r_p$
at each time step $\Delta v$ directly by the instantaneous point-particle
jump $[\Psi_{22}],[\partial_r\Psi_{22}]$ (without using the frequency-domain
snapshots as a boundary condition). Along outgoing null rays $dr/dv=\tfrac12 a_{\rm w}(v,r)$, the
causal propagation time from $r_p\simeq8M_0$ to an extraction sphere $r_{\rm ext}$
is $\Delta v_{\rm prop}(r_p\to r_{\rm ext})=2\int_{r_p}^{r_{\rm ext}}a_{\rm w}^{-1}dr$,
giving $\Delta v_{\rm prop}=92.32M_0$, $154.26M_0$, and $235.92M_0$ for
$r_{\rm ext}=50M_0$, $80M_0$, and $120M_0$, respectively.

Simultaneously, because $\partial_\phi$ is an exact Killing vector of the
spherical accretion metric~\eqref{eq:finite_cloud_metric}, azimuthal angular
momentum balance $\frac{d}{dv}[L(v,r_{\rm orb}(v))]=-\eta\mathcal F_E^{\rm tot}/\Omega$
combined with the exact relativistic circular-orbit thermodynamic identity
$(\partial_r E)_v=\Omega(\partial_r L)_v$ governs the coupled evolution of
the orbital radius $r_{\rm orb}(v)$ and gravitational-wave phase
$\Phi_{\rm GW}^{\rm TD}(v)$:
\begin{align}
 \frac{dr_{\rm orb}}{dv}
 &=-\frac{\eta\,\mathcal F_E^{\rm tot}(v,r_{\rm orb}(v))
 +\Omega(v,r_{\rm orb})\,(\partial_v L)|_{r_{\rm orb}}}
 {(\partial_r E)_v},\nonumber\\
 \frac{d\Phi_{\rm GW}^{\rm TD}}{dv}&=2\Omega(v,r_{\rm orb}(v)),
 \label{eq:coupled_orbital_evolution}
\end{align}
where $(\partial_r E_{\rm vac})|_{r=8M_0}=(r-6M_0)/[2r^3(1-3M_0/r)^{3/2}]
=3.952847\times10^{-3}M_0^{-1}$.

In vacuum at $r_p=8M_0$, the single $(\ell,m)=(2,2)$ polar mode gives
\begin{align}
 \frac{\mathcal F_{22}^{\infty}}{(\mu/M_0)^2}
   &=8.2405141026\times10^{-5},\nonumber\\
 \frac{\mathcal F_{22}^{\rm H}}{(\mu/M_0)^2}
   &=5.4923402966\times10^{-8},
 \label{eq:polar_22_vacuum_number}
\end{align}
so the $m=\pm2$ quadrupole pair contributes
$2\mathcal F_{22}^\infty/(\mu/M_0)^2=1.6481028205\times10^{-4}$, which is
$210.82$ times larger than the $m=\pm1$ axial pair. With the second-order
Frobenius horizon expansion and exact Chandrasekhar--Darboux outer series,
varying the horizon offset between $10^{-4}M_0$ and $10^{-6}M_0$, the outer
matching radius between $60/\omega$ and $140/\omega$, and the asymptotic series
order between 14 and 24 terms changes $\mathcal F_{22}^\infty$ by less than
$8.2\times10^{-12}$ fractionally in vacuum and by $7.9\times10^{-12}$ on the
cloud. At $r_p=400M_0$, the ratio of $\mathcal F_{22}^\infty$ to the leading
Newtonian single-mode flux $(16/5)(\mu/M_0)^2(M_0/r_p)^5$ is $0.98883781$,
agreeing with the analytic relative-1.5PN expansion
$1-\frac{107}{21}(M_0/r_p)+4\pi(M_0/r_p)^{3/2}=0.98883270$ to
$5.11\times10^{-6}$.

\begin{table*}[t]
\caption{Dominant $(\ell,m)=(2,2)$ polar mode solved via the simultaneous
4-component BVP coupled to relativistic EC matter perturbations
($\Phi_{\rm f,22}$), $N=1024$. Columns list the fractional change in
$\mathcal F_{22}^\infty$ relative to vacuum from the simultaneous coupled BVP,
the metric-only fractional change (omitting $\delta n,\delta u^\mu$), and the
excited single-mode matter perturbation flux $\mathcal F_{\rm mat,22}/(\mu/M_0)^2$.}
\label{tab:polar_22_coupled}
\begin{ruledtabular}
\begin{tabular}{@{}cccccc@{}}
$v/M_0$ & Matching & $r_p/M_0$ & Coupled $\Delta\mathcal F_{22}^\infty/\mathcal F_{22}^{\rm vac}$
 & Metric-only $\Delta\mathcal F_{22}^\infty/\mathcal F_{22}^{\rm vac}$ & $\mathcal F_{\rm mat,22}/(\mu/M_0)^2$\\
\hline
0 & Radius & 8.00000000 & $-1.532\times10^{-5}$ & $-1.532\times10^{-5}$ & $6.165\times10^{-12}$\\
0 & Frequency & 7.99997899 & $-1.982\times10^{-6}$ & $-1.982\times10^{-6}$ & $6.165\times10^{-12}$\\
20 & Radius & 8.00000000 & $-7.836\times10^{-6}$ & $-7.836\times10^{-6}$ & $6.131\times10^{-12}$\\
20 & Frequency & 7.99998549 & $1.375\times10^{-6}$ & $1.375\times10^{-6}$ & $6.131\times10^{-12}$\\
\end{tabular}
\end{ruledtabular}
\end{table*}

Table~\ref{tab:polar_22_coupled} compares the simultaneous 4-component BVP
solution $Y_{22}=(\Psi_{22},\Psi'_{22},\Phi_{\rm f,22},\Phi'_{\rm f,22})^T$
against the 2-component metric-only approximation that omits fluid perturbations.
Because the local cloud enthalpy density at $r_p\simeq8M_0$ is scaled by
$\varrho_*M_0^2=10^{-10}$, the two-way fluid-gravity coupling term
$C_{{\rm f}g}\propto\sqrt{16\pi H}$ shifts $\mathcal F_{22}^\infty$ by
$+1.842\times10^{-11}$ fractionally relative to the metric-only value at $v=0$
and excites an acoustic matter flux
$\mathcal F_{\rm mat,22}/(\mu/M_0)^2=6.165\times10^{-12}$
($1.233\times10^{-11}$ for the $m=\pm2$ pair). Comparing the EC spin fluid
($\zeta=10^{-3}$) against the zero-spin control ($\zeta=0$) at fixed initial
$(n,u_r)$ and shared orbital frequency, the EC spin correction reduces the local
sound speed squared at $r_p\simeq8M_0$ from $c_e^2=0.09669$ to $0.08693$
and shifts the coupled $(2,2)$ gravitational luminosity by
$-1.4945\times10^{-8}$ fractionally. Under the boundary and tolerance refinement
check ($\epsilon:10^{-5}M_0\to10^{-6}M_0$, $r_{\rm max}:100/\omega\to140/\omega$,
series terms $20\to24$, $\text{rtol}:10^{-11}\to3\times10^{-12}$), each individual
flux changes by $7.86\times10^{-12}$ ($1.9\times10^3$ times smaller than the
$-1.4945\times10^{-8}$ signal) and the paired EC-minus-zero-spin difference
changes by only $9.05\times10^{-15}$.

\begin{table*}[t]
\caption{Complete 9-mode multipole breakdown ($2\le\ell\le4$, $m=\pm|m|$ pairs
combined) at shared orbital frequency $M_0\Omega=8^{-3/2}$ ($N=1024, v=0$).}
\label{tab:multipole_sum}
\begin{ruledtabular}
\begin{tabular}{cccccc}
Mode $(\ell,\pm m)$ & Parity & Vacuum $\mathcal F_{\ell m}^\infty/(\mu/M_0)^2$
 & Vacuum $\mathcal F_{\ell m}^{\rm H}/(\mu/M_0)^2$
 & Cloud $\Delta\mathcal F_{\ell m}^\infty/\mathcal F_{\ell m}^{\rm vac}$
 & Cloud $\mathcal F_{\ell m}^{\rm mat}/(\mu/M_0)^2$\\
\hline
$(2,\pm2)$ & even & $1.6481\times10^{-4}$ & $1.098\times10^{-7}$ & $-1.982\times10^{-6}$ & $1.233\times10^{-11}$\\
$(2,\pm1)$ & odd & $7.8175\times10^{-7}$ & $1.418\times10^{-8}$ & $5.821\times10^{-6}$ & $0$\\
$(3,\pm3)$ & even & $2.4440\times10^{-5}$ & $8.604\times10^{-10}$ & $-9.930\times10^{-7}$ & $3.627\times10^{-11}$\\
$(3,\pm2)$ & odd & $2.3974\times10^{-7}$ & $1.226\times10^{-10}$ & $8.866\times10^{-6}$ & $0$\\
$(3,\pm1)$ & even & $2.0883\times10^{-9}$ & $5.072\times10^{-11}$ & $-1.269\times10^{-5}$ & $5.195\times10^{-14}$\\
$(4,\pm4)$ & even & $4.5039\times10^{-6}$ & $7.993\times10^{-12}$ & $9.658\times10^{-7}$ & $1.285\times10^{-10}$\\
$(4,\pm3)$ & odd & $5.4572\times10^{-8}$ & $1.018\times10^{-12}$ & $1.194\times10^{-5}$ & $0$\\
$(4,\pm2)$ & even & $2.3948\times10^{-9}$ & $7.712\times10^{-13}$ & $-6.616\times10^{-6}$ & $1.058\times10^{-12}$\\
$(4,\pm1)$ & odd & $7.9715\times10^{-13}$ & $8.634\times10^{-14}$ & $-1.864\times10^{-5}$ & $0$\\
\hline
\shortstack{Total ($2\le\ell\le4$,\\9 pairs)} & Polar+Axial & $1.9483\times10^{-4}$ & $1.251\times10^{-7}$ & $-8.208\times10^{-7}$ & $1.782\times10^{-10}$\\
\end{tabular}
\end{ruledtabular}
\end{table*}

Summing all 9 even- and odd-parity multipole pairs for $2\le\ell\le4$
(Table~\ref{tab:multipole_sum}) yields the total initial EMRI energy flux
$\mathcal F_E^{\rm tot}/(\mu/M_0)^2=1.94959787\times10^{-4}$ at $v=0$
(fractional shift $-8.194\times10^{-7}$ relative to the vacuum total
$1.94959947\times10^{-4}$), with the dominant $(2,\pm2)$ pair supplying
$84.54\%$ of the total luminosity and the four $\ell=4$ pairs contributing
$2.34\%$. Across all five polar pairs in $2\le\ell\le4$, the EC spin term
enhances the total acoustic matter flux by $+2.53\%$ relative to the $\zeta=0$
control ($1.782\times10^{-10}$ versus $1.738\times10^{-10}$).

\begin{figure*}[t]
\centering
\includegraphics[width=.91\textwidth]{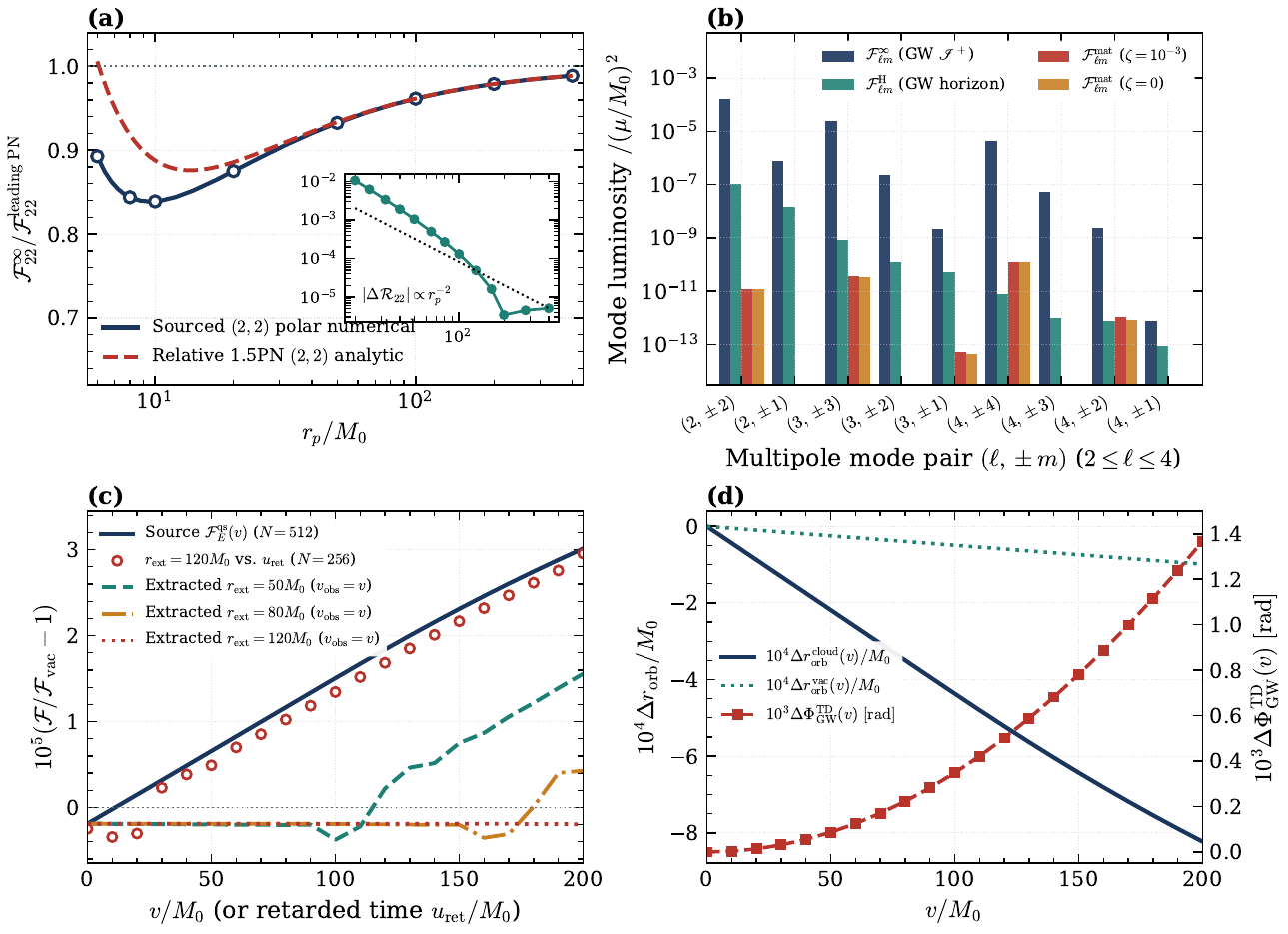}
\caption{Simultaneous 4-component $(\ell,m)=(2,2)$ polar + matter BVP,
complete 9-mode ($2\le\ell\le4$) multipole spectrum, $(1+1)$D PDE
time-domain waveform extraction, and simultaneous coupled orbit-phase evolution.
(a) Vacuum $(2,2)$ luminosity normalized to the leading Newtonian quadrupole
flux versus the analytic relative-1.5PN expansion \cite{Blanchet2014}
(inset: 2PN remainder $\propto r_p^{-2}$).
(b) Complete 9-mode multipole spectrum at $M_0\Omega=8^{-3/2}$ ($N=1024,v=0$)
across all polar and axial pairs for $2\le\ell\le4$.
(c) Time-dependent flux shift $10^5(\mathcal F/\mathcal F_{\rm vac}-1)$
comparing the instantaneous source $\mathcal F_E^{\rm qs}(v)$ against the
$(1+1)$D PDE waveforms extracted at $r_{\rm ext}=50M_0$, $80M_0$, and
$120M_0$ (both at observer time $v_{\rm obs}=v$ and retarded arrival time
$u_{\rm ret}$).
(d) Simultaneous coupled orbital contraction $10^4\Delta r_{\rm orb}(v)/M_0$
(cloud versus vacuum) and coupled time-dependent phase shift
$10^3\Delta\Phi_{\rm GW}^{\rm TD}(v)$ over $v/M_0\in[0,200]$ ($\eta=10^{-5}$).}
\label{fig:polar_coupled}
\end{figure*}

Over the time-dependent background evolution $v\in[0,200M_0]$
($N=512$, Fig.~\ref{fig:polar_coupled}), accretion of the inner cloud increases
the enclosed mass $m(v,r_p)$ from $1.00000157M_0$ to $1.00002408M_0$, driving
the quasi-stationary source flux shift
$\Delta\mathcal F_E^{\rm qs}(v)/\mathcal F_E^{\rm vac}$ at the orbit from
$-1.879\times10^{-6}$ at $v=0$ to $+1.491\times10^{-6}$ at $v=20M_0$,
$+1.506\times10^{-5}$ at $v=100M_0$, and $+3.013\times10^{-5}$ at $v=200M_0$.
Because outgoing null rays emitted from $r_p\simeq8M_0$ at $v_{\rm ret}=0$
require $\Delta v_{\rm prop}=92.32M_0$, $154.26M_0$, and $235.92M_0$ to reach
$r_{\rm ext}=50M_0$, $80M_0$, and $120M_0$ (reaching only $r=102.37M_0$ by
$v=200M_0$), solving the $(1+1)$D Eddington--Finkelstein wave PDE across
$r\in[8M_0,150M_0]$ yields extracted relative flux shifts at $v_{\rm obs}=200M_0$
of $+1.557\times10^{-5}$ at $r_{\rm ext}=50M_0$ (corresponding to retarded
emission around $v_{\rm ret}=107.68M_0$), $+4.286\times10^{-6}$ at
$r_{\rm ext}=80M_0$ ($v_{\rm ret}=45.74M_0$), and $-1.945\times10^{-6}$ at
$r_{\rm ext}=120M_0$ (which lies outside the future null cone of $v_{\rm ret}\ge0$,
$r_p=8M_0$ during $v_{\rm obs}\le200M_0$ and records only the outer-cloud
precursor). Propagating the $(1+1)$D wave PDE across the vacuum exterior to the
retarded arrival window $v_{\rm obs}\in[235.92M_0,435.92M_0]$ at
$r_{\rm ext}=120M_0$ yields an extracted peak flux shift of $+2.956\times10^{-5}$
at $v_{\rm obs}=435.92M_0$ (differing from the source value $+3.013\times10^{-5}$
by $-5.8\times10^{-7}$ due to nonstationary cloud-potential transport, and
agreeing between $N=256$ and $N=512$ to $2.02\times10^{-9}$).

\looseness=-1
Using the exact circular-orbit thermodynamic identity
$\partial_r E(v,r)=\Omega(v,r)\partial_r L(v,r)$
(with $dE_{\rm vac}/dr|_{r=8M_0}=3.952847\times10^{-3}M_0^{-1}$) and
simultaneously integrating $(r_{\rm orb}(v),\Phi_{\rm GW}(v))$ over
$v\in[0,200M_0]$ at $\eta=10^{-5}$, the vacuum reference orbit contracts by
$\Delta r_{\rm vac}(200M_0)=-9.865\times10^{-5}M_0$ while the cloud-embedded
orbit contracts by $\Delta r_{\rm orb}(200M_0)=-8.233\times10^{-4}M_0$
(combining radiation reaction with adiabatic mass-growth contraction
$-\Omega(\partial_v L)/(\partial_r E)$), accumulating a genuine time-dependent
finite-cloud phase shift of
$\Delta\Phi_{\rm GW}^{\rm TD}(200M_0)=+1.367\times10^{-3}\text{ rad}$
while the $(2,\pm2)$ acoustic matter flux decreases from
$1.233\times10^{-11}(\mu/M_0)^2$ to $6.566\times10^{-12}(\mu/M_0)^2$ as the
cloud depletes. As a separate static-background benchmark, integrating across
$r_{\rm vac}\in[6.1M_0,8.0M_0]$ on the frozen $v=0$ profile gives a reference
integral $\Delta\Phi_{\rm GW}^{\rm frozen}=-0.06614\text{ rad}$
($\Delta\Phi_{\rm cons}=-0.53222\text{ rad}$,
$\Delta\Phi_{\rm diss}=+0.46609\text{ rad}$, and
$-0.02424\text{ rad}$ between $\zeta=10^{-3}$ and $\zeta=0$); because the
finite cloud accretes on $v\sim200M_0\ll T_{\rm rad}\sim10^6M_0$, this frozen-slice
integral is strictly a static-slice diagnostic and not a time-dependent phase
prediction for the finite cloud.

\clearpage

\makeatletter
\let\auto@bib@innerbib\@empty
\makeatother

\end{document}